\documentclass[aps,prd,onecolumn,superscriptaddress, nofootinbib, longbibliography, notitlepage]{revtex4-1}

\usepackage{amssymb,amsmath,amsfonts,amsbsy}
\usepackage{color}
\usepackage{enumerate}
\usepackage{graphicx}
\usepackage{pifont}% http://ctan.org/pkg/pifont
\usepackage{appendix}
\DeclareMathAlphabet{\mathpzc}{OT1}{pzc}{m}{it}
\usepackage{aas_macros}
\usepackage{hyperref}
\usepackage{bbm}

\def\d{\mathrm{d}}

\begin{document}

\title{Testing Statistical Isotropy on the Sphere with Minkowski Tensors}

\author{Stephen Appleby}\email{stephen@kias.re.kr}
\affiliation{School of Physics, Korea Institute for Advanced Study, 85
Hoegiro, Dongdaemun-gu, Seoul, 02455, Korea}
\author{Changbom Park} %\email{cbp@kias.re.kr}
\affiliation{School of Physics, Korea Institute for Advanced Study, 85
Hoegiro, Dongdaemun-gu, Seoul, 02455, Korea}

\begin{abstract}
We consider how a class of morphological descriptors, the Minkowski Tensors (MTs), can be used to test the statistical isotropy of random fields on the sphere. The definition of the MTs involves an integral of a tensor, which is an ambiguous operation on a curved manifold. We find that existing estimators in the literature, when applied to fields on the two-sphere, either explicitly break covariance or artificially isotropize the MT due to a geometric rotation of tangent spaces. To evade these issues, we construct the MTs of individual connected components and use them to build correlation functions $\xi_{\pm}(\theta,\nu)$ of their relative orientations. These correlation functions are built from scalars and are therefore covariant, and can be used to search for alignment of structures as a function of scale (angular separation). We generate four sets of random fields on $S^{2}$ -- isotropic, dipole modulated, globally-sheared and locally-sheared, and show how the connected component correlation functions can distinguish scale dependent alignments. For the sheared fields, the decay scale of $\xi_{+}$ measures the angular coherence of the alignment, and the amplitude of $\xi_{-}$ at large separations measures its globally coherent fraction, distinguishing the two shear models. In contrast, dipole modulation generates no significant signal in the orientation statistics. We show analytically that its effect on the traceless component of the Minkowski tensor is second order in the modulation amplitude $\lambda \ll 1$.
\end{abstract}

\maketitle

\section{Introduction} 
\label{sec:introduction}

Testing the statistical isotropy of random fields is a topic of increasing interest to the cosmology community \cite{Aluri:2022hzs,Planck:2019evm,Secrest:2020has,Jones:2023ncn,Battye:2026tsu}. The standard cosmological model is built on the assumption of rotational and translational invariance of the underlying matter distribution, and has proven successful in describing both the background expansion history of the Universe and the growth of density perturbations \cite{Aghanim:2018eyx}. However, a growing body of literature is challenging the notion of statistical isotropy. The temperature fluctuations of the cosmic microwave background exhibit a hemispherical power asymmetry roughly in the direction $(\ell, b) \simeq (225^{\circ} , -27^{\circ})$ \cite{Eriksen:2003db,Hansen:2008ym,Planck:2019evm}, and certain high-redshift quasar and radio source catalogs contain an excess anomalous dipole amplitude \cite{Rubart:2013,Singal:2019,Secrest:2020has}. These signals have motivated the development of increasingly sophisticated statistical methodologies \cite{Hajian:2003qq,Hajian:2005jh,Matsubara:2022ohx,Matsubara:2022eui,Matsubara:2023avg,Matsubara:2024sqn} to characterise existing anomalies and search for associated anisotropies in other data sets, including galaxy catalogs and intrinsic alignments \cite{Inoue:2024twk,Minato:2025ozy}. Thoroughly assessing the significance and physical origin of such signals requires the application of tailored statistics for this purpose.

The Minkowski Functionals are scalar morphological descriptors that characterise the geometry and topology of level sets of random fields. They have a long history of application in cosmology, from analyses of CMB temperature maps to large-scale structure catalogs \citep{1994A&A...288..697M,Schmalzing:1995qn,https://doi.org/10.1111/j.1365-2966.2008.13674.x,2016MNRAS.461.1363N,Chingangbam:2017PhLB,Planck:2019evm,Appleby:2021xoz,Gott:1989yj,1991ApJ...378..457P,Schmalzing:1997aj,Schmalzing:1997uc,1989ApJ...345..618M,1992ApJ...387....1P,Kerscher:1998gs,2001ApJ...553...33P,Park:2009ja,doi:10.1111/j.1365-2966.2010.18015.x,Sahni:1998cr,Bharadwaj:1999jm,2003ApJ...584....1M,vandeWeygaert:2011ip, Park:2013dga,vandeWeygaert:2011hyr,Chingangbam:2013,Shivshankar:2015aza,Pranav:2016gwr,Pranav:2018lox,
Pranav:2018pnu,Feldbrugge:2019tal,Wilding:2020oza,Munshi:2020tzm,Liu:2022vtr,Liu:2023qrj,Liu:2025haj,Rana:2018,Rahman:2021,10.1093/mnras/staf1110}. Being scalar quantities, they are largely insensitive to directional information. The Minkowski Tensors (MTs) are a rank-p generalisation of the Minkowski Functionals \citep{nla.cat-vn2176896, McMullen:1997,Alesker1999,2002LNP...600..238B,HugSchSch07,1367-2630-15-8-083028,JMI:JMI3331,Beisbart:2001gk,Hadwiger, nla.cat-vn1821482,Collischon:2024jhw} that retain directional information, making them useful probes of statistically anisotropic signals \citep{PhysRevE.77.051805,Becker2003ComplexDS,Olszowka2006}. They were first introduced to astrophysical data in \cite{Beisbart:2001vb,2002LNP...600..238B}, and to cosmological random fields in \citep{Ganesan:2017,Chingangbam:2017uqv}. They have subsequently been applied to CMB fields \citep{Chingangbam:2021,K.:2018wpn,Joby:2021,Goyal:2019vkq,Goyal:2021nun}, reionization maps \citep{Kapahtia:2018,Kapahtia:2019ksk,Kapahtia:2021}, and the low-redshift matter density field as traced by galaxies \citep{Appleby:2017uvb,2018ApJ...863..200A}. The ensemble averages of the MTs for anisotropic random fields have been constructed in \cite{2019ApJ...887..128A,Appleby:2022itn,Appleby:2025sbk} and \cite{Klatt_2022} additionally studied higher rank statistics. Numerical algorithms with which to extract the MTs from fields in two and three dimensions can be found in \cite{JMI:JMI3331,Appleby:2017uvb,2018ApJ...863..200A,Schaller2020,Collischon:2024jhw}. Related statistics can be found in Refs. \cite{Kanafi:2023hmr,Abedi:2024ajq,Kanafi:2025gpn}. Recently, \cite{Collischon:2024jhw} introduced irreducible Minkowski tensors on the sphere and a localised analysis via Minkowski Maps, implementing parallel transport to account for the curvature of the two-sphere.

Much of the data that we collect in cosmology is only available as two-dimensional random fields on the sky, either because we only measure a single snapshot at a fixed redshift (the CMB) or we do not have access to accurate three dimensional positional information and so are restricted to projected data. One can study the statistical isotropy of these data, but a subtle issue arises when applying the MTs to fields defined on the two-sphere $S^{2}$. The definition of a Minkowski Tensor involves integrating a tensor-valued quantity over the boundary of an excursion set. On a curved manifold such as $S^{2}$, tensors at different points live in different tangent spaces, and summing them requires a prescription -- such as parallel transport -- to map between those spaces. Without such a prescription, the integral is not covariantly defined. Previous estimators in the literature either implicitly neglect the curvature of the manifold, which yields a non-covariant object, or address it via geodesic parallel transport, which can introduce a systematic bias toward isotropy. Neither outcome is ideal for an unbiased test of statistical isotropy. 

In this work we review these issues. We first show that the standard pixel-sum estimator used to define the Minkowski tensors in two-dimensional flat space ${\mathbb R}^{2}$, does not preserve the tensorial property under local coordinate rotations when applied to fields on $S^{2}$. We then argue that the covariant estimator constructed via geodesic parallel transport systematically suppresses anisotropic signals, limiting its utility as a test statistic. To circumvent both problems, we construct the MTs of individual connected components of the excursion set, and use them to define pairwise orientation correlation functions $\xi_{\pm}(\theta, \nu)$ inspired by weak gravitational lensing statistics. These quantities are built entirely from scalars intrinsic to $S^{2}$, are free from noise-induced bias, and can be used to search for alignment of structures as a function of scale. We validate these statistics on four ensembles of random fields -- isotropic, dipole modulated, globally sheared, and locally sheared -- and study their response to each type of anisotropy. We additionally examine analytically the dipole modulated fields, and explain why they produce no detectable signal in the MTs. We find instead that they generate a signal in the Minkowski functional $W_{1}$, when measured in bands on the sky. 

The paper is organised as follows. In Section \ref{sec:definitions} we review the definition of the Minkowski Tensors in two dimensions. Section \ref{sec:noncov} discusses problems encountered when extracting them from fields on $S^{2}$ and reviews existing estimators. Section \ref{sec:correlation} introduces the connected component correlation functions. Sections \ref{sec:iso} and \ref{sec:results} describe our simulated fields and numerical results respectively. Section \ref{sec:dip} analyzes the dipole modulation case in detail, and we conclude in Section \ref{sec:conclusion}. In the appendix we review the numerical methodology used to extract the MTs from fields defined on the sphere.

\section{Minkowski Tensors in Two Dimensions -- Definitions}
\label{sec:definitions}

In this work we focus on two-dimensional, smooth manifolds ${\cal M}$ over which a mean subtracted field $\delta({\bf x})$ with variance $\sigma^{2} = \langle \delta^2 \rangle$ is defined. We define the excursion set at some threshold \(\nu\) as $Q(\nu) = \{ \mathbf{x} \, | \, \delta(\mathbf{x}) \geq \nu \sigma \}$ with boundary $\partial Q$ set by the curve $\delta({\bf x}) = \nu\sigma$. Typically, $Q$ is not one contiguous region but rather consists of disjoint elements, $Q_{\mu}$, with non-intersecting boundaries $\partial Q_{\mu}$, where $\mu$ indexes each element. Throughout this work we call each individual element a {\it connected component}. 

 From the total excursion set $Q$, we define the Minkowski Tensor:
\begin{equation}\label{eq:def_1} W_{1}^{0,2} = {1 \over A} \int_{\partial Q} \hat{\bf n} \otimes \hat{\bf n} \,  \d \ell \, ,
\end{equation} 
where the integral is over the boundary $\partial Q$, $A$ is the area over which the field is defined, and $\hat{\bf n} = \nabla \delta/|\nabla \delta|$ is the unit normal to $\partial Q$. There exists a larger class of tensors, but in this work we focus on this quantity and refer the reader to \cite{JMI:JMI3331} for a complete accounting. 

It is common to transform the line integral to :
\begin{equation}\label{eq:def_2} W_{1}^{0,2} = {1 \over A} \int_{{\cal M}} \hat{\bf n} \otimes \hat{\bf n} \, \delta_{D}(\delta - \nu\sigma) |\nabla \delta| \, \d A \, ,
\end{equation} 
where $\delta_{D}(x)$ is the Dirac delta function, and we now integrate over the entire manifold ${\cal M}$. Here the delta function picks out points ${\bf x}$ on the excursion set boundary $\delta_{D}(\delta({\bf x}) - \nu\sigma)$, and the tensor product $(\hat{\bf n} \otimes \hat{\bf n}) ({\bf x})$ defines a rank-two tensor in the tangent space $T_{\bf x}{\cal M}$ at each point. On a curved manifold, comparing such tensors requires a prescription such as parallel transport to relate different tangent spaces. Without this, the integral is not covariantly defined. We return to this point in the next section. 

In Euclidean space ${\mathcal M} = {\mathbb R}^{2}$, translation invariance provides a natural identification of tangent spaces at different points, allowing tensors to be averaged with minimal ambiguity. In this case, the integrations in equations (\ref{eq:def_1},\ref{eq:def_2}) can be performed without any issues, because all points can be moved to a common point on the manifold using Euclidean transport. This procedure does not rotate the tangent spaces. If we have a discretized field on a regular Cartesian lattice with basis ${\bf e}_{1}, {\bf e}_{2}$, then we can estimate the Minkowski tensors by replacing the integrals with a pixel sum and discretized derivatives and delta functions 
\begin{equation}\label{eq:def_3} W_{1}^{0,2}|_{i j} =  {1 \over A} \int_{{\mathbb R^{2}}} \delta_{D}(\delta({\bf x}) - \nu\sigma)\,  |\nabla \delta({\bf x})| \, \hat{n}_{i}({\bf x}) \hat{n}_{j}({\bf x}) \d A = {\Delta^{2} \over A} \sum_{m \in {\rm pixels}} \delta_{D}(\delta({\bf x}_{m}) - \nu\sigma) |\nabla \delta({\bf x}_{m})| \hat{n}_{i}({\bf x}_{m}) \hat{n}_{j}({\bf x}_{m})  \, ,
\end{equation} 
where $i,j = 1,2$, we have defined $\hat{\bf n} = n_{1} {\bf e}_{1} + n_{2} {\bf e}_{2}$, $\Delta^{2}$ is the area of a single pixel (assumed equal) and the delta function in the pixel sum is a discretized estimator. This construction relies on the fact that the basis vectors ${\bf e}_{1}$, ${\bf e}_{2}$ are globally constant, so that the components $\hat{n}_{i}$ can be meaningfully summed across different points. From this construction, the eigenvalues of the tensor are typically extracted $\Lambda_{1} \geq \Lambda_{2}$ and a dimensionless measure of anisotropy constructed ; $\alpha = \Lambda_{2}/\Lambda_{1} \leq 1$, with exact isotropy defined as $\alpha=1$ \cite{Ganesan:2017}. Both anisotropic signals in the field, and statistical noise, drive $\alpha$ to lower values. 

The concern of this work is fields defined on $S^{2}$, where the Euclidean description breaks down precisely because the basis vectors become position dependent.

\section{Minkowski Tensors on $S^{2}$}
\label{sec:noncov}

The Minkowski tensors were first generalised to the two-sphere in \cite{Chingangbam:2017uqv}, which introduced the formal definition on curved manifolds and applied them to test the statistical isotropy of random fields \cite{Ganesan:2017,Chingangbam:2017uqv}. That work acknowledged that averaging over distinct structures on $S^{2}$ requires transport to a common point, but the mathematical implementation of this idea was deferred. The formal definition of \cite{Chingangbam:2017uqv} incorporates parallel transport along the curve via tensorial integration, however the estimator adopted there is the flat-space pixel sum (\ref{eq:def_3}) applied to $S^{2}$ : 

\begin{equation}\label{eq:sph_0} W_{1}^{0, 2}|_{ij} = {1 \over A} \int_{S^{2}} \delta_{D}(\delta({\bf x}) - \nu\sigma)\,  |\nabla \delta({\bf x})| \, \hat{n}_{i}({\bf x}) \hat{n}_{j}({\bf x}) \d A \stackrel{?}{=} {\Delta^{2} \over A} \sum_{m \in {\rm pixels}} \delta_{D}(\delta({\bf x}_{m}) - \nu\sigma) |\nabla \delta({\bf x}_{m})| \hat{n}_{i}({\bf x}_{m}) \hat{n}_{j}({\bf x}_{m})  \, ,
\end{equation} 
where $\stackrel{?}{=}$ is used to denote the flat space estimator assumption at the second equality. In the pixel sum after the second equality, $W_{1}^{0, 2}|_{ij}$ is expressed in terms of the components $\hat{n}_{i}({\bf x}_{m})$ of the vector field with respect to a local basis on $S^{2}$. Since these components are defined in different tangent spaces at each pixel location  ${\bf x}_{m}$, a covariant definition of $W_{1}^{0,2}$ requires a prescription to relate them. This is true on any manifold -- on ${\mathbb R}^2$ Euclidean translation is invoked, often implicitly. On $S^{2}$ the analogous canonical choice is parallel transport along great arcs, but this operation is path dependent due to curvature, and so provides no unique identification of tangent spaces. In contrast the pixel sum in (\ref{eq:sph_0}) specifies no prescription, but rather implicitly identifies tangent spaces across the sphere. The resulting estimator is not a tensor, as we now show. Under a local rotation $R$ of the vector components we have $\hat{n}_{i}({\bf x}) \to \hat{n}'_{i}({\bf x}) = R_{i}{}^{k}({\bf x}) \hat{n}_{k}({\bf x})$, which yields the following transformation of the pixel sum in equation (\ref{eq:sph_0}):

\begin{equation}\label{eq:noncov} W_{1}^{0, 2}|'_{ij} \equiv {\Delta^{2} \over A} \sum_{m \in {\rm pixels}} \delta_{D}(\delta({\bf x}_{m}) - \nu\sigma) |\nabla \delta({\bf x}_{m})| R_{i}{}^{k}({\bf x}_{m}) R_{j}{}^{p}({\bf x}_{m}) \hat{n}_{k} \hat{n}_{p}  \, .
\end{equation} 
Since the rotations are pixel-position dependent, they cannot be factored out of the sum, and this makes $W_{1}^{0,2}$ explicitly non-covariant. This can be demonstrated by taking the simple example of two pixels on the sphere $m=1,2$. We define a unit vector at the two points as $\hat{n}({\bf x}_{1}) = (\cos\alpha_{1}, \sin\alpha_{1})$,  $\hat{n}({\bf x}_{2}) = (\cos\alpha_{2}, \sin\alpha_{2})$ and the corresponding MT is $W_{1}^{0,2} = c({\bf x}_{1}) \hat{n}({\bf x}_{1}) \hat{n}({\bf x}_{1}) + c({\bf x}_{2}) \hat{n}({\bf x}_{2}) \hat{n}({\bf x}_{2})$, where $c({\bf x})$ is some unimportant scalar quantity. The trace and determinant of $W_{1}^{0,2}$ are given by 

\begin{equation} {\rm Tr}\left(W_{1}^{0,2}\right) = c({\bf x}_{1}) + c({\bf x}_{2}) \, , \qquad {\rm Det}\left(W_{1}^{0,2}\right) = c({\bf x}_{1}) c({\bf x}_{2}) \sin^{2}(\alpha_{1} - \alpha_{2}) \, .
\end{equation} 
A local coordinate rotation of the vectors can be written as $\hat{n}'({\bf x}_{1}) = (\cos(\alpha_{1}+\theta_{1}), \sin(\alpha_{1}+\theta_{1}))$, $\hat{n}'({\bf x}_{2}) = (\cos(\alpha_{2}+\theta_{2}), \sin(\alpha_{2}+\theta_{2}))$ for arbitrary angles $\theta_{1}$, $\theta_{2}$. We see that the trace is trivially invariant under this rotation, but the determinant is modified as 
\begin{equation}
   {\rm Det}\left(W_{1}^{0,2}\right)' = c({\bf x}_{1}) c({\bf x}_{2}) \sin^{2}(\alpha_{1} - \alpha_{2} + \theta_{1}-\theta_{2})  \, , 
\end{equation}
which implies that the determinant is only invariant under a global rotation $\theta_{1} = \theta_{2}$. The eigenvalues are given by

\begin{equation}\Lambda_{1,2} = {{\rm Tr}\left(W_{1}^{0,2}\right) \pm \sqrt{ \left[{\rm Tr}\left(W_{1}^{0,2}\right)\right]^{2} - 4 {\rm Det}\left(W_{1}^{0,2}\right) } \over 2}\, ,
\end{equation}
and so this simple test shows that the eigenvalues of $W_{1}^{0,2}$, as defined in this section, are not invariant under local coordinate transformations. This issue was partially anticipated in \citep{Chingangbam:2017uqv}, where the requirement of a transport prescription was noted. 

There is one important caveat to this argument : for a statistically isotropic field, we can take the ensemble average  

\begin{eqnarray} \langle W_{1}^{0, 2}|'_{ij} \rangle  &=& {1 \over A} \int_{S^{2}} R_{i}{}^{k}({\bf x}) R_{j}{}^{p}({\bf x})  \big\langle \delta_{D}(\delta - \nu\sigma)\,  |\nabla \delta| \,  \hat{n}_{k}({\bf x}) \hat{n}_{p}({\bf x})\big\rangle  \d A \, , \\
&=& {1 \over A} \int_{S^{2}} R_{i}{}^{k}({\bf x}) R_{j}{}^{p}({\bf x}) \delta_{k p}   \big\langle \delta_{D}(\delta - \nu\sigma)\, F(|\nabla \delta|)\big\rangle   \d A \, ,\\
&=&  {\mathbbm{1}_{ij} \over A} \int_{S^{2}}  \big\langle \delta_{D}(\delta - \nu\sigma)\, F(|\nabla \delta|)\big\rangle   \d A  \, ,
\end{eqnarray} 
where $\mathbbm{1}_{ij}$ is the Kronecker delta and $F(|\nabla \delta|)$ is some unimportant function of the scalar quantity $|\nabla \delta|$. We see that statistical isotropy forces all directions to be equivalent, so $\langle W_{1}^{0, 2}|_{ij}\rangle$ is rotationally invariant and the tensorial transformation property is restored. This holds only in expectation, statistical fluctuations break exact isotropy\footnote{We note that in \citep{Chingangbam:2017uqv} it was stated that the off-diagonal elements of the ensemble-averaged tensor vanish for Gaussian fields irrespective of any departure from isotropy. However, this holds only when the principal axes of the anisotropy are aligned with the coordinate basis, otherwise off-diagonal elements are non-zero \cite{Chingangbam:2021}.}. 

The quantity defined in equation (\ref{eq:sph_0}) and introduced in \cite{Chingangbam:2017uqv} can be unambiguously extracted from a random field defined on $S^{2}$, and can be sensitive to certain anisotropic signals if the random field $\delta({\bf x})$ is statistically anisotropic. However, this sensitivity cannot be relied upon as evidence of either the presence or absence of a genuine signal. We show in Appendix \ref{app:stereographic} that a different choice of frame, obtained by stereographically projecting a field and measuring the tensor eigenvalues in Cartesian coordinates on the plane, can null a real anisotropic signal (cf. Figure \ref{fig:stereo}, bottom left panel). Since the choice of local frame on $S^{2}$ is arbitrary, this ambiguity cannot be resolved from a single non-covariant measurement. We conclude that the estimator (\ref{eq:sph_0}), computed in any single fixed frame using a non-covariant pixel sum, should not be used in isolation as a test of statistical isotropy. 

We note, however, that in \cite{K.:2018wpn,Goyal:2019vkq,2020arXiv200615038K} the non-covariant form was used in a frequentist comparison between data and isotropic simulations computed in the same fixed frame. Such a like-for-like test retains a well-defined significance level, and any detected departure from the isotropic ensemble is meaningful within that frame. The sensitivity to a specific anisotropic signal will nevertheless depend on the choice of frame (cf. Appendix C). Analyses of patches that are small relative to the curvature scale of the sphere will also be less affected by these ambiguities \cite{Goyal:2021nun,Bashir:2025gen}.

A minor, separate issue is that the ratio of eigenvalues $\alpha$ is equal to unity only in the limit of exact statistical isotropy, which is not realised for finite-area fields \cite{Chingangbam:2021}. Statistical fluctuations, set by the underlying power spectrum, will generically bias the measured value such that $\langle \alpha \rangle < 1$. Disentangling statistical fluctuations from intrinsic anisotropy therefore requires knowledge of the statistical properties of the field (e.g. its power spectrum, non-Gaussian corrections), introducing additional model dependence. 

Finally, we comment on the trace ${\rm Tr} (W_{1}^{0,2})$, which should be a scalar and equal to the Minkowski Functional $W_{1}$. One might question whether either of these statements remain true for fields defined on $S^{2}$, given that $W_{1}^{0,2}$, as defined in this section, does not transform as a tensor. However, even for our non-covariant definition, the condition $W_{1} = {\rm Tr} (W_{1}^{0,2})$ holds. The reason is that the trace operator is linear, so it commutes with the integration over $S^{2}$. This implies that we can calculate the trace of $W_{1}^{0,2}$ at every point on the manifold, generating a set of scalars which can then be summed irrespective of any parallel transport choice. This is in contrast to the eigenvalues $\Lambda_{1}$, $\Lambda_{2}$ of $W_{1}^{0,2}$, used to define $\alpha$. These quantities are formed from non-linear combinations of the tensor components, so the act of transforming tensor components to eigenvalues does not commute with the integration over $S^{2}$. This explains why the trace of $W_{1}^{0,2}$ remains equal to the scalar $W_{1}$ regardless of how we construct it, but the eigenvalues will not be invariant under local coordinate rotations in general.

\subsection{Covariant Estimator}
\label{sec:cov}

In \cite{Appleby:2022itn}, the authors addressed the covariance issue in the Minkowski tensor estimator when applied to three-dimensional large scale structure data in redshift space\footnote{Covariance on $S^{2}$ was also addressed in \cite{Collischon:2024jhw}, where parallel transport to a common point was used to define irreducible Minkowski tensors.}. Although the real space matter density field is assumed to be statistically isotropic in ${\mathbb R}^{3}$, the redshift space distortion operator is radial relative to an observer located at some ${\bf r} = 0$, and it generates a distinct line-of-sight anisotropy in the radial direction defined by the unit basis vector ${\bf e}_{r}$. The Minkowski tensors of the observed matter field are therefore naturally described using spherical coordinate systems defined on the manifold $S^{2} \times {\mathbb R}_{>0}$. The question addressed in that work was how to construct an estimator that preserved the tensorial structure of the MTs, while simultaneously extracting the radial anisotropy generated by redshift space distortion. 

Since \cite{Appleby:2022itn} considered three-dimensional fields and used slightly cumbersome notation, for simplicity we write down the two-dimensional analogue of their estimator, applicable for fields on $S^{2}$. We define $P_{\gamma}({\bf x}_{m} \to {\bf x}_{m_{0}})$ as parallel transport along great arc $\gamma$ from ${\bf x}_{m}$ to ${\bf x}_{m_{0}}$,\footnote{In \cite{Appleby:2022itn} we used shorthand notation ${}^{\gamma}\nabla_{i}\delta$ to describe the same operation of parallel transport along great arcs. For clarity, here we are more explicit but the method is identical.} and the Minkowski tensor can be expressed as 

\begin{equation} \label{eq:cov_1} W_{1}^{0,2}|_{ij} \stackrel{?}{=} {\Delta^{2} \over A} \sum_{m \in {\rm pixels}} \delta_{D}(\delta({\bf x}_{m})-\nu\sigma) {\left[P_{\gamma}\nabla\delta({\bf x}_{m})\right]_{i}\left[P_{\gamma}\nabla\delta({\bf x}_{m})\right]_{j} \over |\nabla \delta({\bf x}_{m})|} \, ,
\end{equation}

\noindent where we have again used the symbol $\stackrel{?}{=}$ to highlight the ambiguity in the definition. In this case, the $\stackrel{?}{=}$ symbol reflects the ambiguity in the choice of path $\gamma$ and position ${\bf x}_{m_{0}}$ -- different paths yield different tensors at ${\bf x}_{m_{0}}$. This expression is also a sum over pixels, but now the vector $\nabla\delta({\bf x}_{m})$ at ${\bf x}_{m}$ is transported to a common location ${\bf x}_{m} \to {\bf x}_{m_0}$ on the manifold via parallel transport along great arcs, prior to taking the average. Transport corresponds to a path-dependent rotation of vectors between tangent spaces, generating a well-defined linear map between them. Furthermore, because all vectors are transported to the same tangent space, they now transform under a single common rotation at $m_{0}$. Specifically, $\left[P_{\gamma}\nabla\delta({\bf x}_{m})\right]_{i}$ are components of a vector in the tangent space $T_{{\bf x}_{m_{0}}}S^{2}$. A change of basis at $m_{0}$ via a single rotation $R({\bf x}_{m_{0}})$ acts on all transported vectors identically, yielding the standard transformation law $W_{1}^{0,2}|'_{ij} = R_{i}{}^{k}({\bf x}_{m_{0}})R_{j}{}^{p}({\bf x}_{m_{0}}) \, W_{1}^{0,2}|_{kp}$. This is in contrast to the ${\mathbb R}^{2}$ estimator (cf. equation (\ref{eq:sph_0})) where position-dependent rotations prevent such a factorization. This estimator realizes the averaging methodology first proposed in \citep{Chingangbam:2017uqv}. If the field is statistically isotropic on $S^{2}$, then the point $m_{0}$ at which the average is taken is irrelevant in terms of expectation values. If the field is anisotropic, then the choice of $m_{0}$ likely matters and is an additional ambiguity.  

By applying the operator $P_{\gamma}$, we have made two choices -- parallel transport and great arc pathing. The Minkowski tensor components are sensitive to both choices. If $\gamma$ represents a path different from the great arc, then the sum in (\ref{eq:cov_1}) will yield a different outcome. This path dependence reflects the non-zero curvature of $S^{2}$, and is generic to curved manifolds. Given that any path apart from great arcs introduces additional structure on the sphere, in this work we do not consider any other choice. Parallel transport is similarly selected as the unique transport rule that introduces no additional structure beyond the path itself.

Even with these minimal choices, the act of parallel transport will rotate the vectors from ${\bf x}_{m}$ to ${\bf x}_{m_{0}}$. The transported vector's orientation at ${\bf x}_{m_{0}}$ reflects both the geometry of the path and the local field direction at ${\bf x}_{m}$. When summed over pixels, the path-dependent rotations are distributed across all directions on $S^{2}$, partially averaging out coherent preferred direction in the field. This implies that the proper, covariant definition of the tensor will suppress preferred directions and bias the result toward isotropy. In \cite{Appleby:2022itn}, statistical isotropy was assumed on $S^{2}$ and covariance was the primary concern; in the present work we are testing for anisotropy, and the isotropization bias of geodesic transport becomes an important systematic\footnote{One might construct some designer anisotropic field for which the geometric rotation actually reinforces the anisotropic signal in the field. We do not claim that all fields are isotropized, but constructing a counter example would likely require significant contrivance.}.

In the left panel of Figure \ref{fig:schem} we present an example of the geometric rotation. We define two unit vectors $\hat{n}_{A}$ and $\hat{n}_{B}$ at points $A$ and $B$ respectively. Both are pointing north. We transport both vectors to a new location $m_{0}$ along the great arcs $\gamma_{A}$, $\gamma_{B}$. After this operation, neither transported vector $P_{\gamma}\hat{n}_{A}$ nor $P_{\gamma}\hat{n}_{B}$ points north at $m_{0}$, and the two are not parallel $P_{\gamma}\hat{n}_{A} \, . \, P_{\gamma}\hat{n}_{B} \neq 1$. Note that if we transported $\hat{n}_{A}$ along the meridian to $B$, and then transported both $\hat{n}_{A}$ and $\hat{n}_{B}$ from $B$ to $m_{0}$ along $\gamma_{B}$, then they would be parallel at $m_{0}$ but not pointing north. This highlights the path dependence of the result. In this work we only use single great arc paths such as $\gamma_{A}$ and $\gamma_{B}$ for transport.

To summarize this section : a direct sum of tensor elements at different locations on $S^{2}$ yields a non-covariant object with uncertain geometric meaning. Defining a tensor via covariant averaging necessarily requires rotating vectors and generates a result that is path dependent and can be biased towards isotropy. Both outcomes are problematic. The goal of this work is to use the Minkowski Tensors to construct a covariant statistic that can be used to unambiguously test for statistical isotropy of random fields on the sphere. Fortunately, such mathematical machinery has already been developed in the field of weak lensing.

\section{Connected Component Correlation Functions}
\label{sec:correlation}

As expressed in Section \ref{sec:definitions}, the excursion set $Q$ is composed of a set of distinct connected components labeled $Q_{\mu}$ with non-intersecting boundaries $\partial Q_{\mu}$ and areas $A_{\mu}$. For each connected component we can define its individual Minkowski Tensor 

\begin{equation}
 \label{eq:com_1}   W_{1, \mu}^{0,2} =  {1 \over A_{\mu}} \int_{\partial Q_{\mu}} \hat{\bf n} \otimes \hat{\bf n} \,  \d \ell \, ,
\end{equation}
and from this construct its eigenvalues $\Lambda_{1, \mu} \geq \Lambda_{2, \mu}$, corresponding eigenvectors ${\bf v}_{\mu}$, ${\bf u}_{\mu}$ and dimensionless shape parameter $\beta_{\mu} = \Lambda_{2, \mu}/\Lambda_{1, \mu} \leq 1$. In Appendix \ref{app:numerics_I} we review the numerical methodology used to extract these statistics from a random field on $S^{2}$. In what follows, we only use the eigenvector associated with the principal axis of the connected component, ${\bf v}_{\mu}$. 

For each $\mu$ connected component, we find its geometric center ${\bf \bar{p}}_{\mu} = \sum_{\eta \in \mu} {\bf p}_{\eta}$, normalised to a unit vector, on $S^{2}$, and great-arc, geodesic transport the vectors ${\bf \hat{n}}$ to ${\bf \bar{p}}_{\mu}$ before evaluating the sum in equation (\ref{eq:com_1}). Even this short, local transport from the boundary to ${\bf \bar{p}}_{\mu}$ will slightly distort and isotropize the individual connected component morphologies. We call this the `finite-size geometric distortion' and attempt to quantify its significance in the Appendix. 

\begin{figure}
    \centering
    \includegraphics[width=0.98\textwidth]{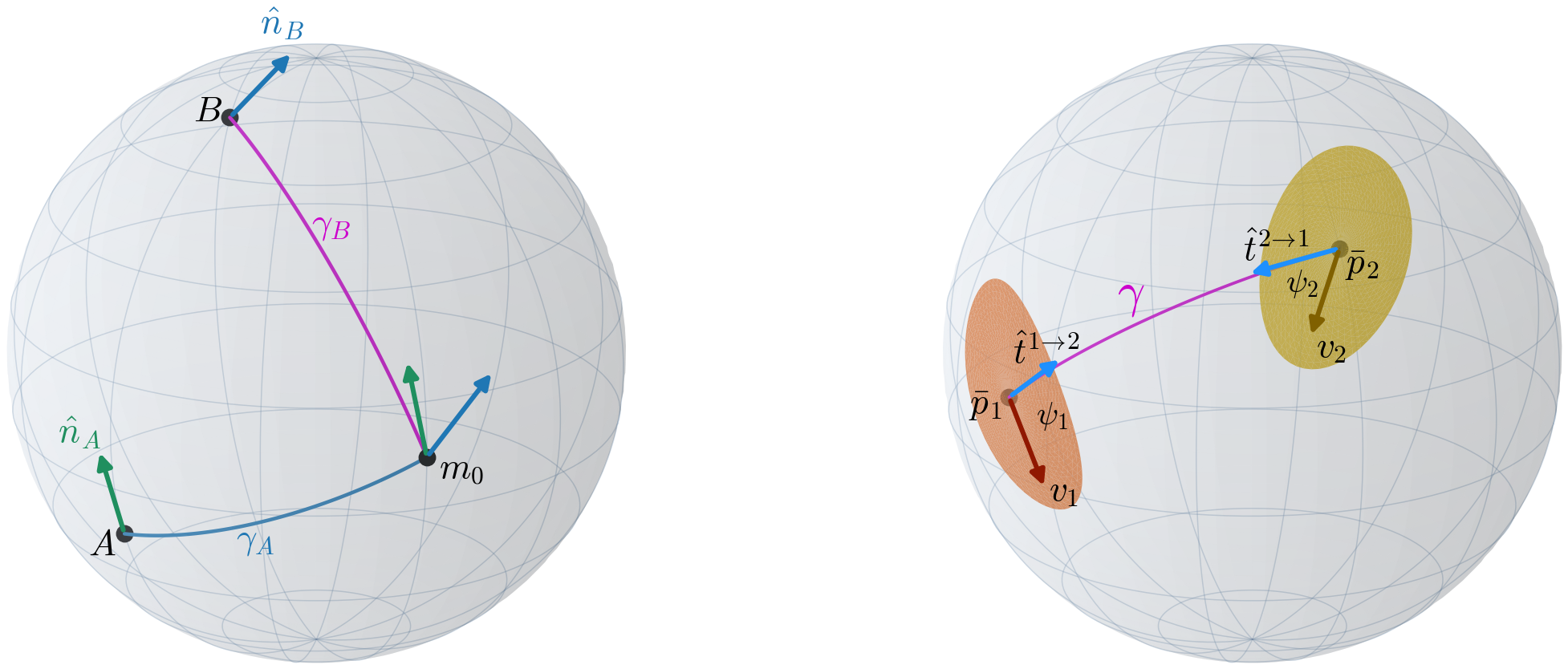} 
\caption{[Left Panel] An example of the geometric rotation of vectors generated by the covariant averaging procedure outlined in Section \ref{sec:cov}. Points $A$ and $B$ lie on a common meridian, and the unit vectors $\hat{n}_{A}$ and $\hat{n}_{B}$ are both pointing north. After geodesic transport to $m_{0}$ along $\gamma_{A}$ and $\gamma_{B}$ respectively, neither $P_{\gamma}\hat{n}_{A}$ nor $P_{\gamma}\hat{n}_{B}$ is pointing north. [Right panel] A schematic image of two connected components on $S^{2}$ (orange/yellow ellipses), the great arc joining them (purple curve) and the relevant angles and vectors defined at ${\bf \bar{p}}_{1}$ and ${\bf \bar{p}}_{2}$ used in the definition of $\xi_{\pm}$. The excursion sets are presented here as ellipses, but  will be irregular shapes for random fields.}
    \label{fig:schem}
\end{figure}

At each point ${\bf \bar{p}}_{\mu}$, we construct the Minkowski tensor $W_{1,\mu}^{0,2}$, and extract its eigenvalue ratio $\beta_{\mu}$ and eigenvector of the largest eigenvalue ${\bf v}_{\mu}$. We also define the direction ${\bf \hat{t}}^{(\mu \to \eta)} \in T_{\bar{p}_{\mu}}S^{2}$ as the tangent vector of the geodesic from ${\bf \bar{p}}_{\mu}$ to a second connected component defined at ${\bf \bar{p}}_{\eta}$, and similarly the geodesic tangent vector pointing to ${\bf \bar{p}}_{\mu}$ at ${\bf \bar{p}}_{\eta}$ as ${\bf \hat{t}}^{(\eta \to \mu)} \in T_{\bar{p}_{\eta}}S^{2}$. These vectors lie in the respective tangent planes of their connected component locations, and specifically live in different vector spaces. The angles $\psi_{\mu} = \angle ({\bf v}_{\mu}, {\bf \hat{t}}^{(\mu \to \eta)})$ and $\psi_{\eta} = \angle ({\bf v}_{\eta}, {\bf \hat{t}}^{(\eta \to \mu)})$ are well defined scalars independent of any coordinate choice on $S^{2}$, and no transport is required to define them. In the right panel of Figure \ref{fig:schem} we present a schematic of the geometric quantities introduced in this section. 

Finally, we define
\begin{equation} \epsilon_{+} = - |\epsilon|\cos 2\psi \, , \quad \epsilon_{\times} = - |\epsilon|\sin 2\psi  \, ,
\end{equation} 
with $\epsilon = (1 - \beta)/(1+\beta)$ and use these to build the correlation functions 
\begin{eqnarray} & & \xi_{+}(\theta,\nu) = \big\langle \epsilon_{+}^{(\mu)} \epsilon_{+}^{(\eta)} + \epsilon_{\times}^{(\mu)} \epsilon_{\times}^{(\eta)} \big\rangle \, , \\
& & \xi_{-}(\theta,\nu) = \big\langle \epsilon_{+}^{(\mu)} \epsilon_{+}^{(\eta)} - \epsilon_{\times}^{(\mu)} \epsilon_{\times}^{(\eta)} \big\rangle \, ,
\end{eqnarray} 
where $\theta$ is the shortest great arc distance between points ${\bf \bar{p}}_{\mu}$ and ${\bf \bar{p}}_{\eta}$ and the angular brackets denote a pair-wise average over the connected components. We note that this definition could be extended to study the correlation between alignment of connected components at different thresholds $\nu$ and $\nu'$. The weight $|\epsilon| \geq 0$ is adopted because it naturally down-weights connected components close to circular $\beta \simeq 1$ which carry little information about preferred directions. 

These correlation functions measure, in a coordinate independent manner, the average degree of alignment between connected components defined on $S^{2}$ as a function of their separation, and are lifted directly from the weak lensing literature \cite{1991ApJ...380....1M,Kaiser:1991qi,Schneider:2002jd,Bartelmann:1999yn}. In weak lensing analysis, the object studied is the shear field $\gamma$, a spin-2 quantity describing the coherent distortion of background galaxy shapes. Although the specific tensor differs between this work and weak lensing, the analysis proceeds analogously.

The quantities defined in this section have a number of useful properties. First, they are practically (i.e. up to the finite-size geometric distortion) covariant -- all quantities used to construct $\xi_{\pm}(\theta,\nu)$ are scalars defined using only intrinsic structures on $S^{2}$, specifically great arcs. No path-dependent geometric rotations are required because tensor components in distinct tangent spaces are not directly averaged. Second, the commonly used anisotropy parameter $\alpha = \Lambda_{2}/\Lambda_{1} \leq 1$ is strictly bounded from above via its definition, and therefore is subject to noise induced bias $\langle \alpha \rangle < 1$ even for statistically isotropic fields. In contrast, the correlation functions $\xi_{\pm}(\theta,\nu)$ are not bounded quantities, so statistical fluctuations in finite-area realisations scatter the measured values symmetrically about the ensemble expectation. For a statistically isotropic field, isotropy requires $\xi_{\pm}$ to depend only on the pair separation $\theta$, and the decay of field correlations beyond the smoothing scale forces $\xi_{\pm}(\theta) \to 0$ for $\theta \gg \theta_{G}$. At small separations, isotropic fields generically exhibit non-zero correlation due to local alignment of neighbouring excursion sets. Coherent alignment persisting at separations far beyond the correlation length constitutes a signal of statistical anisotropy, and the scale dependence of $\xi_{\pm}(\theta,\nu)$ characterizes the angular coherence of the alignment. In Section \ref{sec:results} we consider how various anisotropic fields will imprint a signal on $\xi_{\pm}(\theta, \nu)$.

\section{Isotropic Gaussian Random Fields}
\label{sec:iso}

Before studying particular anisotropic configurations, we first construct a set of statistically isotropic fields on $S^{2}$ as a baseline test of our analysis pipeline. The details of our numerical generation of random fields, and the subsequent extraction of the Minkowski functionals and tensors, is described in the Appendix. In this section we provide a short summary.

Prior to defining the fields, we clarify our normalisation conventions. For a statistically isotropic field the ensemble average $\langle \delta^{2} \rangle = \sigma_{0}^{2}$ is position independent, and equals the spatial average $\bar{\sigma}^{2}$ via ergodicity. For an anisotropic field, the local ensemble average $\langle \delta^{2} \rangle = \sigma^{2}(\theta,\phi)$ is position dependent and generally distinct from both the spatial average $\bar{\sigma}^{2}$, and the variance of the isotropic component $\sigma_{0}^{2}$. Numerically we always analyze fields normalised by the spatial average $\delta/\bar{\sigma}$, which for the isotropic case is equivalent to the unit normalised quantity $\delta/\sigma_{0}$.

We place $N_{\rm p}$ points randomly and isotropically distributed on the sphere, assigning each point angles $\phi_{i} = 2\pi u_{i}$, $\theta_{i} = \cos^{-1}(2v_{i} - 1)$ where $u_{i},v_{i} \sim {\cal U}(0,1]$ are uniformly distributed. Each point has corresponding unit vector ${\bf p}_{i} = (\sin\theta_{i}\cos\phi_{i}, \sin\theta_{i}\sin\phi_{i},\cos\theta_{i})$ in $\mathbb{R}^{3}$. The spherical Delaunay triangulation of these points is then generated using a convex hull algorithm. 

The points are dressed with a Gaussian random field using the expression 

\begin{equation}
\delta_{i} = \sum_{\ell=1}^{\ell_{\rm max}} \left[ a_{\ell 0} A_{\ell 0} P_{\ell 0}(\cos\theta_i) + \sum_{m=1}^{\ell} \sqrt{2} A_{\ell m} P_{\ell
  m}(\cos\theta_i) \left[ a_{\ell m} \cos (m\phi_{i}) + b_{\ell m} \sin (m\phi_{i}) \right] \right]
  \end{equation}
The sum begins at $\ell=1$, so $\delta$ is zero mean by construction with variance $\sigma_{0}^{2} = \langle \delta^{2} \rangle$. We introduce a power spectrum $C_{\ell} = \ell^{n} \exp[-\ell (\ell+1) \theta_{G}^{2}]$ including an angular smoothing $\theta_{G}$ and we fix $n=0$. The smoothing is selected to be three times the mean particle separation to suppress shot noise, $\theta_{G} = 3 \bar{\theta}$ with $\bar{\theta} = \sqrt{4\pi/N_{p}}$. Then we define $a_{\ell m} = \sqrt{C_{\ell}} x$, $b_{\ell m} = \sqrt{C_{\ell}} y$, where $x,y$ are unit variance Gaussian distributed random variables, and use the $a_{\ell m}$, $b_{\ell m}$ random variables to fix $\delta_{i}$ at each point. $A_{\ell m}$ is a normalisation factor defined in (\ref{eq:norm}) and we select $\ell_{\rm max} = 192$. This value includes all relevant modes not suppressed by the Gaussian smoothing for our choice of $N_{p}$. 

Finally, the Minkowski functionals, Betti numbers and tensors are constructed using a marching triangle algorithm, described in Appendix \ref{app:numerics_I}. Individual connected components are found by performing a graph search through the triangles, and the location of the $\mu$ connected component is defined as an average over the points ${\bf p}_{\eta}$ belonging to that object ${\bf \bar{p}}_{\mu} = \sum_{\eta \in \mu} {\bf p}_{\eta}$, normalised to a unit vector. The eigenvalues and eigenvectors of the Minkowski tensor $W_{1,\mu}^{0,2}$ are defined at ${\bf \bar{p}}_{\mu}$, and the pairwise correlation functions $\xi_{\pm}(\theta,\nu)$ are constructed.

 \begin{figure}
    \centering
    \includegraphics[width=0.7\textwidth]{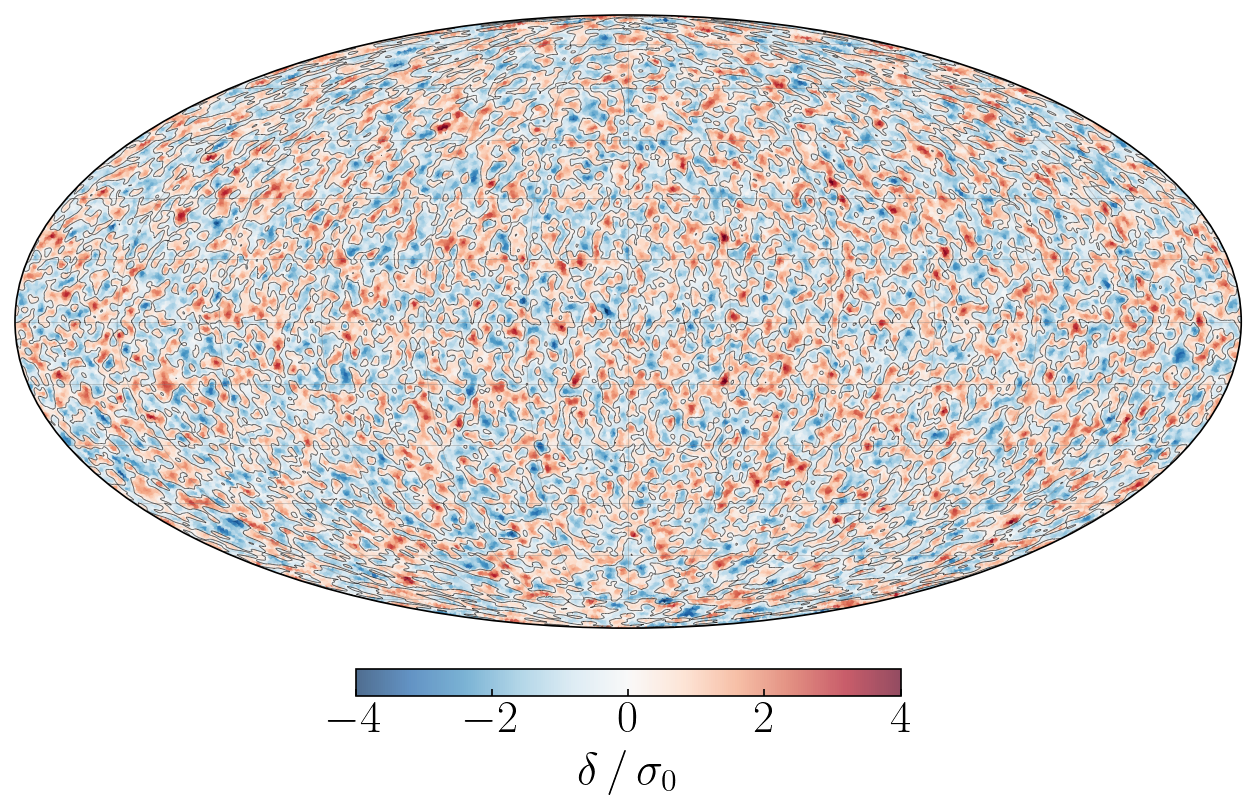} 
\caption{Example of a Gaussian random field $\delta/\sigma_{0}$ defined on $S^{2}$, drawn from a flat power spectrum and smoothed with scale $\theta_{G} = 0.7 \, {\rm deg}$. The iso-field line $\delta = 0$, generated using the marching triangle algorithm presented in the Appendix, is shown as a faint black solid set of curves. }
    \label{fig:1}
\end{figure}

\begin{figure}
    \centering
    \includegraphics[width=0.99\textwidth]{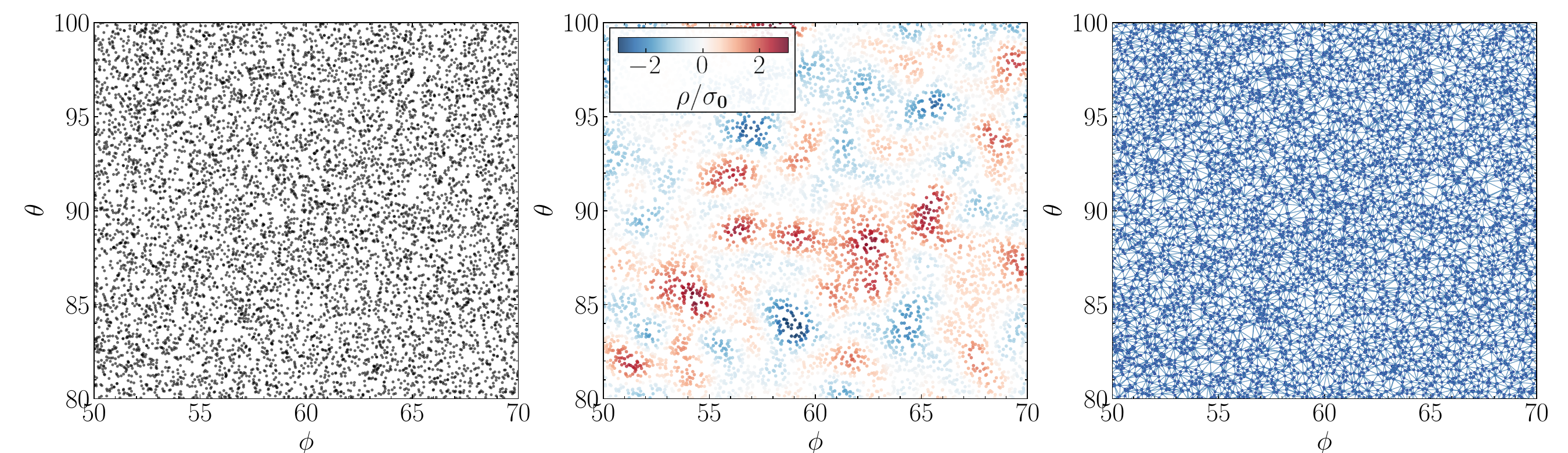} 
\caption{A $20 \times 20 \, {\rm deg}^{2}$ patch of the density field presented in Figure \ref{fig:1}. In the left panel we present the isotropic points in the patch, in the central panel we present the same points dressed with their density values, and in the right panel the spherical Delaunay triangulation used for the marching triangle algorithm.}
    \label{fig:2}
\end{figure}

In the top left panel of Figure \ref{fig:1} we present a Mollweide projected density field constructed using our algorithm. This corresponds to a Gaussian random field with a flat power spectrum $n = 0$ and smoothed on scale $\theta_{G} = 3\times\sqrt{4\pi/N_{p}} \, {\rm rad}$ (chosen simply as a few times the mean point separation). We use $N_{p} = 800,000$ points isotropically distributed on the sphere to define the field and triangulation, which yields $\theta_{G} \simeq 0.7 \, {\rm deg}$. A small, non-projected $20 \times 20 {\rm deg}^{2}$ patch of the field is presented in Figure \ref{fig:2}. Specifically, in the left panel we present the points in this patch, in the central panel the points are dressed with their density values, and in the right panel we present the triangulated mesh used for the marching triangle algorithm. In the full Mollweide projection of the density field, we present the closed contours for the $\nu=0$ threshold, constructed using the marching triangle algorithm described in Appendix \ref{app:numerics_I}. 

We generate $N_{\rm real} = 400$ realisations of a Gaussian random field, for each we construct the Minkowski tensors and correlation functions $\xi_{\pm}(\theta,\nu)$. To test the response of the Minkowski Tensors to statistical anisotropy, we also generate three anisotropic random fields which we now describe.

\subsection{Anisotropic Random Fields}
\label{sec:anisotropy}

In addition to the isotropic realisations, we generate three distinct anisotropic fields on $S^{2}$ and study the response of the correlation functions to each type. We list the anisotropic models below. Our base model is the isotropic field defined in the previous section, which we denote as $\delta_{\rm iso}$ in what follows. For each case, the degree of anisotropy is determined by a single free parameter that we fix to $\lambda=0.76$ to ensure a large $\sim 40\%$ dipole modulation amplitude and a $\sim 4\%$ shear contribution in what follows. At each point on the sphere we define an orthonormal tangent basis $\{{\bf e}_{\theta},{\bf e}_{\phi}\}$, where ${\bf e}_{\theta}$ is chosen to point along the great circle toward the north pole, and ${\bf e}_{\phi}$ is the orthogonal azimuthal direction.

\begin{enumerate}
    \item Dipole modulation :  
    \begin{equation} \label{eq:ani1} \delta = \delta_{\rm iso} \left( 1 + \lambda Y_{10}(\theta,\phi) \right) \, ,
    \end{equation} 
    with direction chosen such that the field has more power in the north pole relative to the south. Isotropy is broken by coupling $\ell$ and $\ell \pm 1$ modes. This is analogous to the CMB hemispherical power asymmetry.  
    \item Global shear :
    \begin{equation} 
      \delta = \delta_{\rm iso} + \lambda \, \bar{\theta}^{2}  Q_{ab}  H^{ab} \, ,
    \end{equation}
    where $a,b = (\theta,\phi)$, repeated indices are summed and we introduce the traceless, symmetric spin-2 tensor that generates a preferred direction in the tangent plane at every point on $S^{2}$ : 
    \begin{equation}
     \label{eq:glob_proj}   Q = {1 \over 2} \begin{pmatrix}
1 & 0 \\
0 & -1
\end{pmatrix} \, ,
    \end{equation}
    and the Hessian is 
    \begin{equation}\label{eq:hess} H_{ab} =  \nabla_{a}\nabla_{b} \delta_{\rm iso} \, ,
    \end{equation} 
    where here and throughout this work we define $\nabla_{\phi} = (1/\sin\theta)\partial_{\phi}$. The choice of $Q_{ab}$ generates a global anisotropy with a preferred axis pointing away from the north pole at every point on the sky (along meridians), so the 
    alignment is coherent over the entire sphere. 
    \item Local shear :
    \begin{equation} 
      \delta = \delta_{\rm iso} + \lambda \, \bar{\theta}^{2} \tilde{Q}_{ab}  H^{ab} \, .
    \end{equation}
    The Hessian $H_{ab} = \nabla_{a}\nabla_{b} \delta_{\rm iso} $ is the same as the global shear case, but now an independent, large-scale Gaussian random field $\tilde{\delta}$ is drawn with a power spectrum restricted to $C_{\ell=4} = C_{\ell=5} \neq 0$ and all other modes zero. At each point, the local gradient of $\tilde{\delta}$ defines a preferred direction 
    \begin{equation}
    \hat{{\bf d}} = \frac{\nabla\tilde{\delta}(\mathbf{x})}{|\nabla\tilde{\delta}(\mathbf{x})|} 
    \, .
\end{equation}
The spin-2 tensor is then constructed as:
    \begin{equation}
     \label{eq:loc_proj}   \tilde{Q} = \begin{pmatrix}
\hat{d}_{\theta}^{2} - 1/2 & \hat{d}_{\theta}\hat{d}_{\phi} \\
\hat{d}_{\theta}\hat{d}_{\phi} & \hat{d}_{\phi}^{2} - 1/2
\end{pmatrix} \, .
    \end{equation}
The vector $\hat{\bf d}$ is randomly directed on $S^{2}$ but correlated over scales $\sim 180^{\circ}/\ell$, generating local alignment of structures. We note that $\tilde{Q}$ is quadratic in $\hat{\bf d}$ and hence invariant under $\hat{\bf d} \to -\hat{\bf d}$; this rectification transfers power to low, even multipoles of the alignment pattern irrespective of the band limit imposed on $\tilde{\delta}$. In the limiting case $\tilde{\delta} \propto Y_{10}$, the direction field is $\hat{\bf d} = -{\bf e}_{\theta}$ everywhere and the construction reduces exactly to the global shear model (\ref{eq:glob_proj}). The locally sheared field therefore contains a sub-dominant, globally coherent alignment component in addition to its dominant finite-range structure.
\end{enumerate}

%%%%%%%%%

The second and third cases introduce a shear-type anisotropy analogous to the tidal fields measured in weak lensing analysis, which also couples a spin-2 preferred-direction tensor to the local curvature of the field. We use these designer fields purely to test the response of the statistics $\xi_{\pm}(\theta,\nu)$ to different types of anisotropy. 

\begin{figure}
    \centering
    \includegraphics[width=0.98\textwidth]{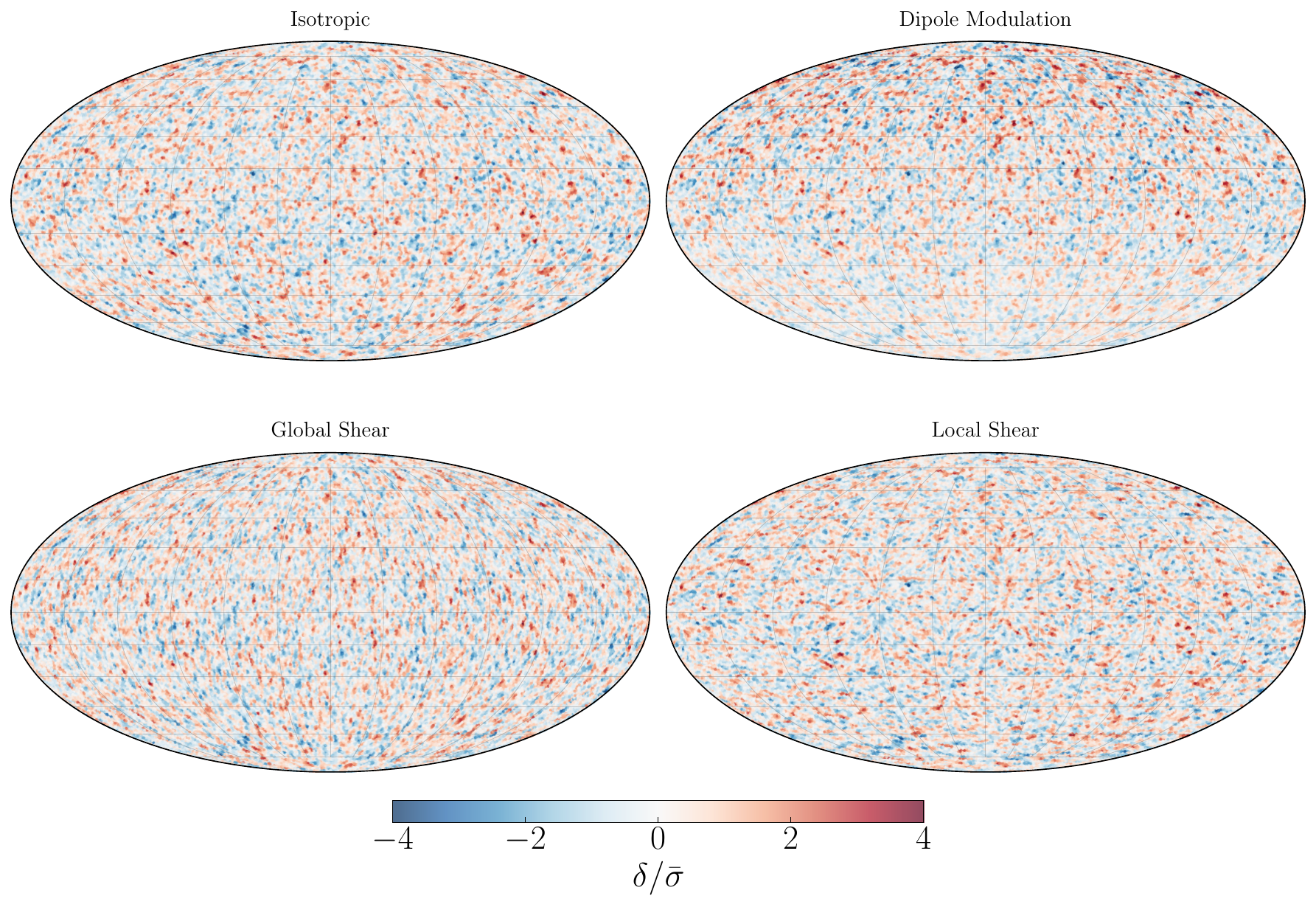} 
\caption{An example of a single realisation of the four fields generated in this work : Isotropic (top left), dipole modulated (top right), global shear (lower left) and local shear (lower right). For the dipole modulated field we have fixed $\lambda = 0.76$, but for the lower panels we have taken an unreasonably large shear amplitude $\lambda = 10$ to make the shear pattern distinguishable.}
    \label{fig:fields}
\end{figure}

Once we have generated the density fields, we unit variance normalise them using the spatial average, generating the fields $\delta \to \delta/\bar{\sigma}$. In Figure \ref{fig:fields} we present a single realisation of $\delta_{\rm iso}$ (top left panel) and the same realisation after the three anisotropic operations have been applied to it : dipole modulation (top right panel), global (bottom left) and local (bottom right) shear. For the dipole modulation field, we have fixed $\lambda = 0.76$, but for the shear examples in the bottom panels we have taken $\lambda = 10$, to greatly exaggerate the anisotropic contribution for the sake of visual clarity. 

Before proceeding, we comment on the typical size of the anisotropy for the three cases. For the dipole modulation, the parameter $\lambda$ entirely determines the magnitude of the anisotropy. For the shear models, the variance of the isotropic and shear fluctuations are given by
\begin{eqnarray} & &  \sigma_{\rm iso}^{2} = {1 \over 4\pi} \sum_{\ell} (2\ell+1) C_{\ell} = \sigma_{0}^{2} \, , \\
& & \sigma^{2}_{\rm shear} = \frac{\lambda^{2}\bar{\theta}^{4}}{32\pi} \sum_{\ell} (2\ell+1)\,\ell^{2}(\ell+1)^{2} C_{\ell} = \frac{\lambda^{2}\bar{\theta}^{4}\sigma_{2}^{2}}{8}\, ,
\end{eqnarray}
so the typical size of the shear contribution relative to the isotropic fluctuations is of order $\sigma_{\rm shear}/\sigma_{\rm iso} = \lambda\bar{\theta}^{2}\sigma_{2}/(\sqrt{8}\,\sigma_{0})$, where $\sigma_{2}^{2}$ is the standard isotropic cumulant defined in equation (\ref{eq:sig2}). For our choice of a flat power spectrum and Gaussian smoothing $\theta_{G} = 3\bar{\theta}$, the fractional contribution of the shear anisotropy is $\simeq 0.056\lambda$, that is approximately $4\%$ for our choice $\lambda = 0.76$.

\section{Results}
\label{sec:results}

\begin{figure}
    \centering
    \includegraphics[width=0.98\textwidth]{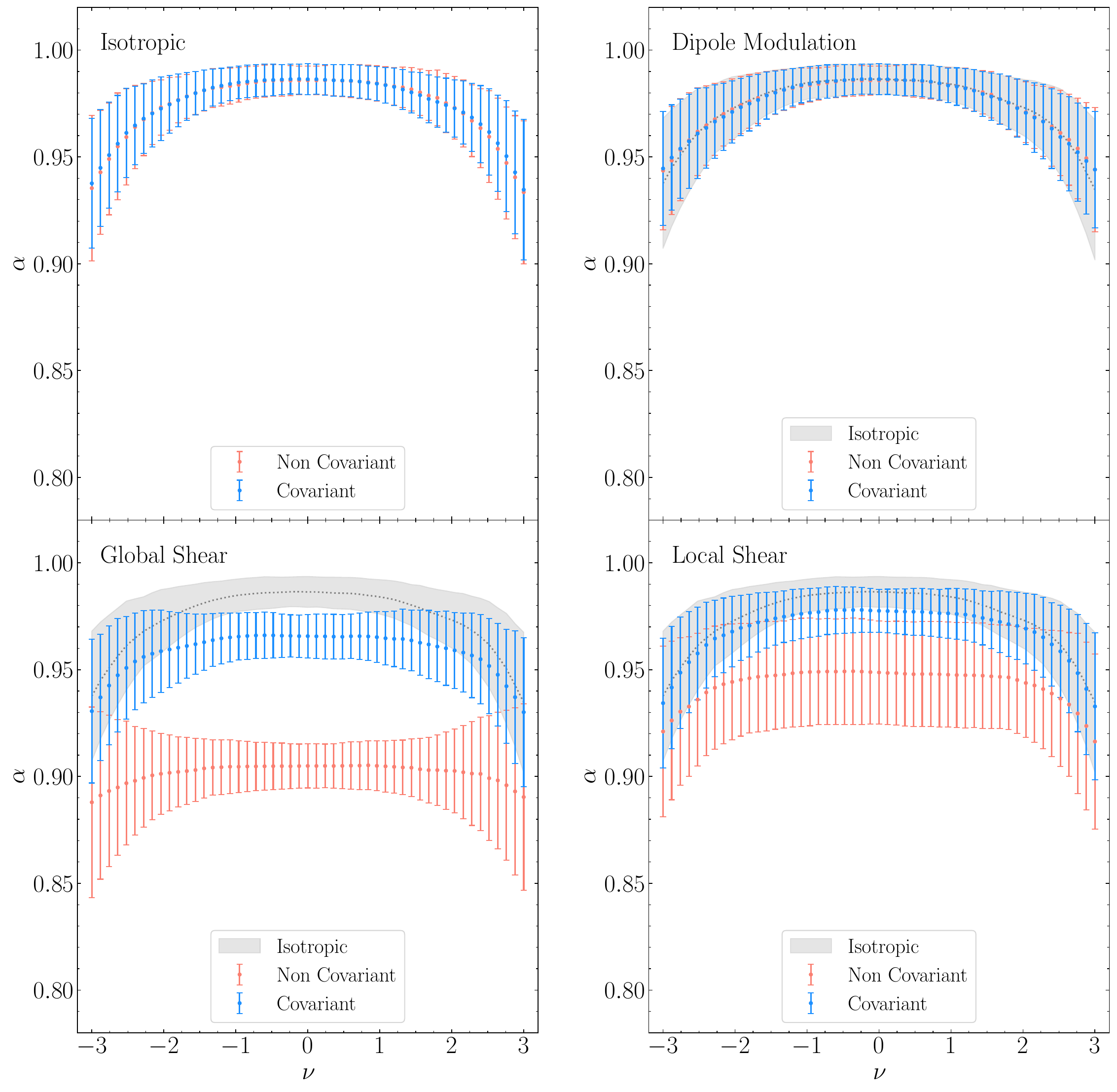} 
\caption{The statistic $\alpha$ extracted from $N_{\rm real}=400$ realisations of an isotropic random field (top left), dipole modulated (top right), globally sheared (bottom left) and locally sheared (bottom right). The points/error bars are the median and $16/84\%$ ranges of the realisations, and the red/blue points correspond to the non-covariant/covariant estimators described in Sections \ref{sec:noncov} and \ref{sec:cov} respectively. The grey filled area is the isotropic result from the top left, reproduced in each panel for reference.  }
    \label{fig:alpha}
\end{figure}

In Figure \ref{fig:alpha} we present the standard one-point statistic $\alpha = \Lambda_{2}/\Lambda_{1} \leq 1$ that is used to test for anisotropic signals \cite{Ganesan:2017,Chingangbam:2017uqv}, for the four fields considered in this work. Specifically, we present the median and $16/84\%$ limits on $\alpha$ extracted from $N_{\rm real}=400$ realisations of an isotropic (top left), dipole modulated (top right), global shear (bottom left) and local shear (bottom right) fields. The red points/error bars are measurements of the non-covariant estimator described in Section \ref{sec:noncov} and the blue points/error bars the covariant form of the pixel sum defined in Section \ref{sec:cov}, with fiducial point chosen arbitrarily as ${\bf x}_{m_{0}} = (60^{\circ}, 45^{\circ})$. In the three anisotropic panels, we present the isotropic result as a filled grey region for reference. 

For the isotropic fields, both covariant and non-covariant estimators yield statistically identical results, as expected. Both exhibit some signal with $\alpha < 1$, but this corresponds to realisation scatter for finite area data. In the top right panel, we find that $\alpha$ is insensitive to the presence of dipole modulation, even for a large $\sim 40\%$ anisotropic amplitude scaling. We consider this case in more detail in section \ref{sec:dip}, because it has direct relevance for modern cosmological data sets in the form of the CMB power asymmetry and the quasar anomalous dipole \cite{Planck:2019evm,Secrest:2020has}. Conversely, in the lower panels, we observe that $\alpha$ is sensitive to both types of shear anisotropy. There is a signal in the covariant (blue) and non-covariant (red) estimators, with both showing a systematically lower $\alpha$ value across all thresholds compared to the isotropic realisations (grey filled). The covariant estimator is closer to unity, reflecting the bias towards isotropy that is a result of the covariant averaging procedure. The globally sheared fields (bottom left panel) show a larger signal than the locally sheared case (bottom right panel), as expected. Although the non-covariant version of $\alpha$ inherits a strong signal from the sheared fields, we show in Appendix \ref{app:stereographic} that the signal can be entirely washed out by a different choice of coordinate system. 

\begin{figure}
    \centering
    \includegraphics[width=0.98\textwidth]{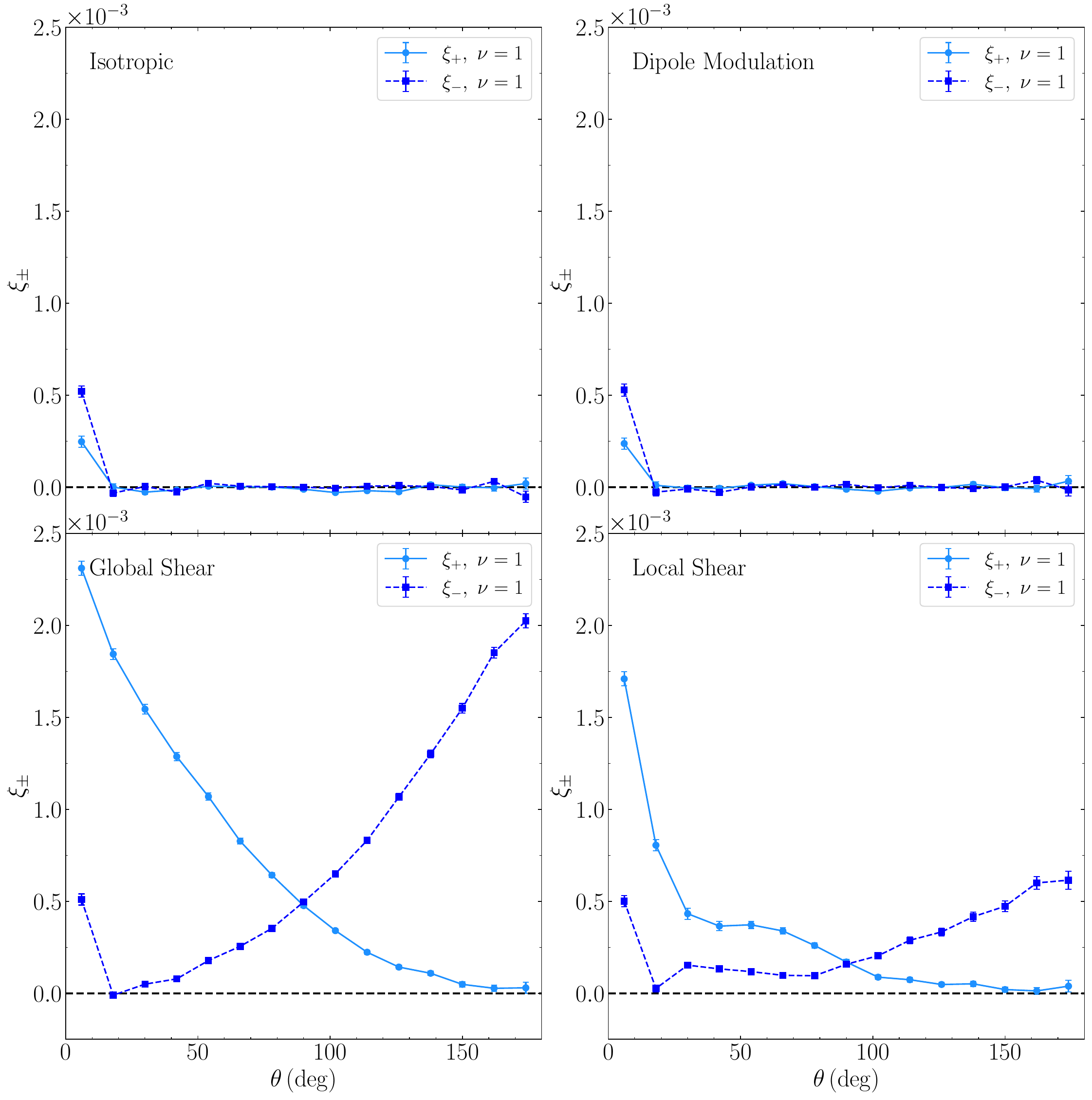} 
\caption{The connected component correlation functions $\xi_{+}(\theta)$ (pale solid blue line/error bars) and $\xi_{-}(\theta)$ (dark dashed blue line/error bars)  as a function of $\theta$ for threshold $\nu=1$. The panels represent the four ensembles of fields presented in Figure \ref{fig:alpha}. Points/error bars are the mean and standard error of $N_{\rm real}=400$ realisations.}
    \label{fig:corr}
\end{figure}

In Figure \ref{fig:corr} we present the correlation functions $\xi_{+}(\theta)$ and $\xi_{-}(\theta)$, for the four field ensembles. We present the mean and error on mean of $N_{\rm real}=400$ simulations as points/error bars. We only consider the $\nu = 1$ threshold as the curves are largely similar for other thresholds, and the pale solid/dark dashed blue lines correspond to $\xi_{+}(\theta)$ and $\xi_{-}(\theta)$ respectively. We use $N_{\theta} = 15$ angular bins equi-spaced over the range $0 \leq \theta \leq 180^{\circ}$. 

\begin{figure}
    \centering
    \includegraphics[width=0.98\textwidth]{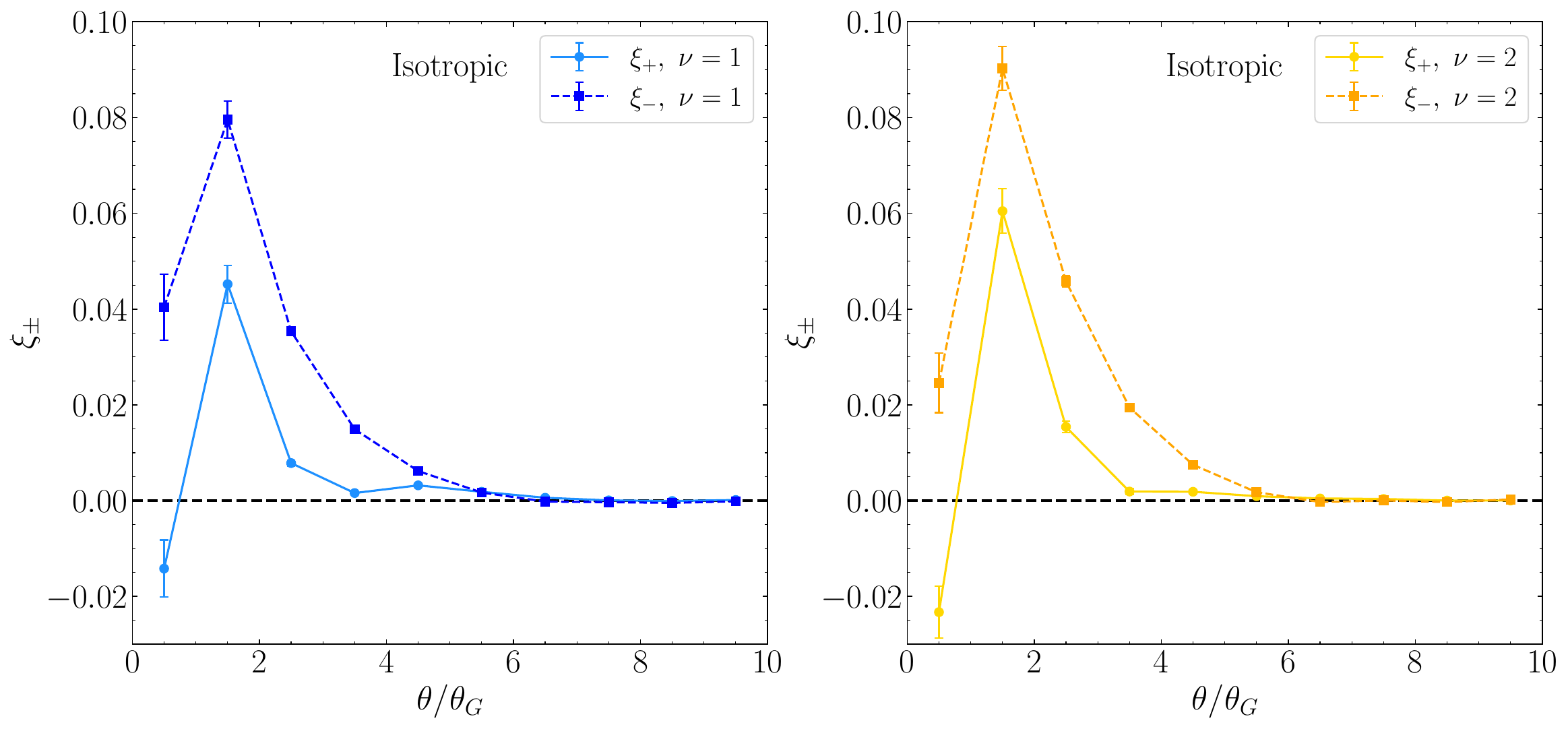} 
\caption{The connected component correlation functions $\xi_{+}(\theta)$ and $\xi_{-}(\theta)$ as a function of $\theta/\theta_{G}$ for two threshold values $\nu = 1$ (left panel), $\nu=2$ (right panel), in ten fine bins over the range $0 \leq \theta \leq 10\theta_{G}$. This figure resolves the first angle bin in Figure \ref{fig:corr}. We present the isotropic ensemble only; the dipole modulated ensemble is identical, and the sheared ensembles exhibit the same short-range structure superposed on their additional alignment signals. Points/error bars are the mean and standard error of $N_{\rm real}=400$ realisations.}
    \label{fig:small_corr}
\end{figure}

First, we observe a small excess correlation in both $\xi_{\pm}$ in the smallest angular bin, for all four field ensembles. In Figure \ref{fig:small_corr} we resolve this bin into 10 finer bins over the range $0 \leq \theta \leq 10 \theta_{G}$ for the isotropic ensemble. The left/right panels are $\nu=1$/$\nu=2$ respectively, and the light solid/dark dashed curves are $\xi_{+}(\theta)$ and $\xi_{-}(\theta)$. At the smallest separations $\theta < \theta_{G}$, the two functions have opposite signs, with $\xi_{+}$ negative. This combination identifies a specific geometry: for a pair whose principal axes are tilted at equal and opposite angles $\psi_{\mu} = -\psi_{\eta} \equiv \bar{\psi}$ relative to the connecting geodesic, the pair contributes $\cos 4\bar{\psi}$ to $\xi_{+}$ but $+1$ to $\xi_{-}$. Mirror-symmetric configurations therefore always generate positive $\xi_{-}$, whereas $\xi_{+}$ turns negative once the tilt exceeds $22.5^{\circ}$. Such configurations arise from a selection effect. Components separated by less than the smoothing scale can remain distinct only if the intervening saddle point lies well away from the connecting geodesic, since an on-axis saddle would require the field to fall below then re-rise above the threshold over less than a smoothing length. Both components elongate toward the common off-axis saddle, producing large opposite tilts. The measured ratios $\xi_{+}/\xi_{-} \simeq -0.35$ ($\nu = 1$) and $\simeq -0.95$ ($\nu = 2$) correspond to typical tilts $\bar{\psi} \simeq 27^{\circ}$ and $\bar{\psi} \simeq 40^{\circ}$ respectively -- higher-threshold pairs that survive at these separations are increasingly contorted, with the $\nu=2$ value consistent with the maximal mirror configuration $\bar{\psi} = 45^{\circ}$. At separations $\theta \gtrsim \theta_{G}$ the intervening saddle typically lies close to the connecting geodesic, the principal axes point approximately toward each other, and both $\xi_{\pm}$ become positive, peaking at $\theta \simeq 1.5 \theta_{G}$. The excess of $\xi_{-}$ over $\xi_{+}$ at the peak reflects a residual off-axis displacement. Beyond $\theta \sim 5 \theta_{G}$ the field decorrelates and both functions are consistent with zero. In the top panels of Figure \ref{fig:corr}, there is no statistically significant correlation in any other angular bin for the isotropic or dipole modulated fields.

For the globally sheared field (lower left panel, Figure \ref{fig:corr}), the correlation function $\xi_{+}(\theta)$ is large and positive at small separations, and decays monotonically with increasing $\theta$. At small separations the behaviour is intuitive: all connected components are elongated along their local meridians, and over scales small compared to the curvature of $S^{2}$ the geodesic connecting a pair makes approximately the same angle with the meridian at both endpoints, $\psi_{\mu} \simeq \psi_{\eta}$. Every pair therefore contributes positively to $\xi_{+} \sim \langle \cos 2(\psi_{\mu} - \psi_{\eta}) \rangle$, regardless of the orientation of the pair relative to the alignment axis. The decay of $\xi_{+}$ with separation is a purely geometric effect. The field remains equally anisotropic at all points on the sphere, but a great arc does not maintain a constant angle with the meridians it crosses -- the connecting geodesic rotates relative to the alignment pattern as the separation grows, and the angles $\psi_{\mu} \neq \psi_{\eta}$ increasingly differ. This is in contrast to the flat space analogue: a straight line crossing a uniformly aligned pattern in $\mathbb{R}^{2}$ subtends a constant angle with the alignment axis at all points, and in flat space $\xi_{+}$ would exhibit no decay.

The signal is not lost at large separations, but shifts into $\xi_{-}(\theta)$. As the pair separation approaches the antipodal limit $\theta \to 180^{\circ}$, the geometry of great arcs on $S^{2}$ drives the endpoint angles towards equal and opposite values, $\psi_{\eta} \to -\psi_{\mu}$ -- the mirror configuration to which $\xi_{-}$ is sensitive. The falling $\xi_{+}$ and rising $\xi_{-}$ in the lower left panel are therefore a single coherent alignment signal, understood through a reference geodesic that rotates with respect to the alignment pattern. For this alignment pattern, which is symmetric under the antipodal map, the two functions satisfy the mirror relation $\xi_{-}(\theta) \simeq \xi_{+}(180^{\circ} - \theta)$, which can be verified directly in the figure: the amplitude of $\xi_{-}$ in the final angular bin matches that of $\xi_{+}$ in the first bin (up to the short-range intrinsic contribution, which is not part of the coherent pattern and has no mirror image), and the two curves cross at $\theta = 90^{\circ}$. We stress that the migration of the signal from $\xi_{+}$ to $\xi_{-}$ is a property of the statistic on the curved manifold, not of the field: the anisotropic alignment is equally coherent at all separations, and the pair $\xi_{\pm}(\theta)$ must be interpreted jointly when searching for alignment signals over large fractions of the sky.

The locally sheared field (lower right panel of Figure \ref{fig:corr}) presents very different phenomenology. The correlation function $\xi_{+}(\theta)$ is large and positive at small separations -- the first angular bin also contains the short-range intrinsic contribution common to all four ensembles -- but decays rapidly, falling by an order of magnitude within $\theta \sim 30^{\circ}$, in contrast to the slow geometric decay of the global case. This rapid drop-off reflects the finite coherence of the alignment direction $\hat{\bf d}$, and is the central discriminating feature between the two shear models: the correlation functions directly resolve the angular scale over which structures are coherently aligned. 

The detailed shape of the panel contains two further features, both of which originate in the construction of the large scale direction-field $\hat{\bf d} = \nabla \tilde{\delta}/|\nabla \tilde{\delta}|$ rather than in the statistics themselves. First, $\xi_{+}$ does not fall directly to zero but exhibits a shoulder extending to $\theta \sim 50^{\circ}$. The normalisation of the gradient removes the amplitude of $\tilde{\delta}$, and the coherence of the remaining pure direction field is set not by the correlation length of $\tilde{\delta}$ but by its critical points, at which $\hat{\bf d}$ is undefined. For power restricted to $\ell = 4,5$ there are only a few tens of such points on the sphere, partitioning the manifold into domains of size $\sim 40^{\circ}$ within which the direction field varies smoothly. The alignment coherence therefore persists over the domain scale, beyond the correlation length of $\tilde{\delta}$. Second, $\xi_{-}$ exhibits a weak rise at large separations, reaching $\sim 30\%$ of the small-scale $\xi_{+}$ amplitude at the antipode. This is the same geometric migration observed in the global shear panel, sourced by a sub-dominant coherent component of the alignment pattern, generated by the construction itself. Specifically, the tensor $\tilde{Q}$ is quadratic in $\hat{\bf d}$, and therefore transfers power from the $\ell=4,5$ input into the lower multipoles. The `locally' sheared field therefore carries a sub-dominant, globally coherent alignment, which migrates from $\xi_{+}$ into $\xi_{-}$ precisely as in the lower left panel, but with correspondingly suppressed amplitude -- note that the $\xi_{\pm}$ curves cross at the same $\theta \simeq 90^{\circ}$ in both panels, the signature of the even-parity symmetry that the quadratic construction enforces. 

Taken together, the two shear panels demonstrate the joint discriminating power of the correlation functions: the decay scale of $\xi_{+}$ measures the angular coherence of the alignment, while the large-separation behaviour of $\xi_{-}$ measures the globally coherent fraction of the alignment pattern. A field with purely finite-range alignment would generate no antipodal $\xi_{-}$ signal, whereas fully coherent alignment migrates completely, as in the global shear case.

\section{Dipole Anisotropy} 
\label{sec:dip}

One of the interesting byproducts of our analysis is the fact that the dipole modulated anisotropy, characterised by equation (\ref{eq:ani1}), does not generate a detectable signal in the Minkowski Tensors -- neither $\alpha$ nor $\xi_{\pm}$ (cf. top right panels, Figures \ref{fig:alpha}-\ref{fig:corr}). This implies that dipole power modulation does not generate alignment between excursion sets even when the dipole modulation is very large $\lambda \sim {\cal O}(1)$. In this section we explore this case in more detail, because it is the most relevant form of anisotropy that appears in both the CMB power asymmetry and large scale structure dipole studies \cite{Planck:2019evm,Secrest:2020has}. In this section we use the non-covariant definition of the tensor, to avoid the additional complication of isotropization due to transport. This case was previously studied numerically in \cite{Goyal:2019vkq}.

First, in this work we are taking $\delta_{\rm iso}$ to be a Gaussian random field, which means that $\delta$ is also Gaussian, but multiplied by a deterministic, position-dependent amplitude shift $[1 + \lambda Y_{1 0}(\theta)]$. Such a dipole modulation generates a non-stationary field on $S^{2}$. This implies that when we take the ensemble average of the Minkowski tensor, the operation does commute with the area integral due to linearity, but the area integral does not collapse to $4\pi$ copies of the ensemble average ; 

\begin{equation} \langle W_{1}^{0,2}|_{ij} \rangle = {1 \over 4\pi} \int_{S^{2}} \Big\langle {\nabla_{i} \delta \nabla_{j}\delta \over |\nabla \delta|} \delta_{D}(\delta - \nu) \Big\rangle \d A \neq \Big\langle {\nabla_{i} \delta \nabla_{j}\delta \over |\nabla \delta|} \delta_{D}(\delta - \nu) \Big\rangle \, .
\end{equation} 
The second consequence of dealing with a non-stationary field is that the field and its gradient are correlated at $x$ : $\langle \delta(x) \nabla \delta(x) \rangle \neq 0$. This generates additional correlation structure that modifies the shape of the Minkowski functionals and tensors as a function of threshold $\nu$ \cite{Appleby:2022itn}.

Starting from the density field, we calculate its mean and variance at some $\theta,\phi$. In our example, there is no $\phi$ dependence because we have selected the dipole to be aligned with the poles of our coordinate system : 

\begin{equation} \langle \delta \rangle = \langle \delta_{\rm iso} \rangle \left[  1 + \lambda \sqrt{3 \over 4\pi} \cos\theta  \right] = 0, \qquad  \langle \delta^{2} \rangle = \langle \delta_{\rm iso}^{2} \rangle \left[  1 + \lambda \sqrt{3 \over 4\pi} \cos\theta  \right]^{2} \, .
\end{equation}
In our analysis we normalise the field by the angle averaged variance defined over $S^{2}$, which is $\bar{\sigma}^{2} = \sigma_{0}^{2} (1 + \lambda^{2}/4\pi)$. So, the field that we will study is $ \delta/\bar{\sigma}$. The gradients are given by 
\begin{eqnarray} & & \delta_{\theta} \equiv \nabla_{\theta} \delta = {1 \over \bar{\sigma}}  \left[ \left( 1 + \lambda \sqrt{3 \over 4\pi} \cos\theta \right) \nabla_{\theta}\delta_{\rm iso} - \lambda \sqrt{3 \over 4\pi} \sin\theta \delta_{\rm iso}  \right] \, , \\ 
& &  \delta_{\phi} \equiv \nabla_{\phi}\delta = {1 \over \bar{\sigma}} \left( 1 + \lambda \sqrt{3 \over 4\pi} \cos\theta \right) \nabla_{\phi}\delta_{\rm iso}  \, ,
\end{eqnarray}
where $\nabla_{\phi} = (1/\sin\theta)\partial_{\phi}$, and the relevant non-zero, one-point covariances are 
\begin{eqnarray}  \sigma^{2}(\theta) &\equiv&  \langle \delta^{2} \rangle = {1 \over 1 + \lambda^{2}/4\pi}\left[ 1 + \lambda \sqrt{3 \over 4\pi} \cos\theta  \right]^{2}   \, , \\
 \sigma_{\theta}^{2}(\theta) &\equiv& \langle \delta_{\theta}^{2} \rangle =  {1 \over 1 + \lambda^{2}/4\pi}\left[ \left( 1 + \lambda \sqrt{3 \over 4\pi} \cos\theta \right)^{2} {\sigma_{1}^{2} \over 2\sigma_{0}^{2}}  + {3 \lambda^{2} \over 4\pi}  \sin^{2}\theta  \right]  \, , \\
  \sigma_{\phi}^{2}(\theta)  &\equiv& \langle \delta_{\phi}^{2} \rangle = {1 \over 1 + \lambda^{2}/4\pi} \left( 1 + \lambda \sqrt{ 3\over 4\pi}  \cos\theta \right)^{2} {\sigma_{1}^{2} \over 2\sigma_{0}^{2}}  \, , \\ 
  \sigma_{\times}^{2}(\theta) &\equiv&  \langle \delta \delta_{\theta} \rangle =   - {\lambda \over 1 + \lambda^{2}/4\pi}  \sqrt{3 \over 4\pi} \sin\theta \left( 1 + \lambda \sqrt{3 \over 4\pi} \cos\theta \right) \, ,
\end{eqnarray}
where we have defined the isotropic cumulants as 
\begin{eqnarray} & & \sigma_{0}^{2} = {1 \over 4\pi} \sum_{\ell} (2\ell+1) C_{\ell} \, , \\
& & \sigma_{1}^{2} = {1 \over 4\pi} \sum_{\ell} (2\ell+1) \ell (\ell+1) C_{\ell} \, , \\
\label{eq:sig2} & & \sigma_{2}^{2} = {1 \over 4\pi} \sum_{\ell} (2\ell+1)\ell^{2}(\ell+1)^{2}C_{\ell} \, .
\end{eqnarray} 
The two most important points are that $\langle \delta_{\theta}^{2} \rangle \neq \langle \delta_{\phi}^{2} \rangle$ and $\langle \delta \delta_{\theta} \rangle \neq 0$. Note that we have absorbed the factor $1/\sin\theta$, from the covariant derivative, into the definition of the random field $\delta_{\phi}$. This eliminates the explicit $\theta$ dependence from the covariance matrix and $|\nabla \delta| = \sqrt{\delta_{\theta}^{2} + \delta_{\phi}^{2}}$ for the unmodulated field ($\lambda=0$). To calculate the Minkowski tensor, we use the joint Gaussian PDF 
\begin{equation} P(\delta, \delta_{\theta}, \delta_{\phi}) = {1 \over \sqrt{(2\pi)^{3}|C|}} \exp \left[-X C^{-1} X^{\rm T}/2 \right] \, ,
\end{equation}
where $X = (\delta, \delta_{\theta}, \delta_{\phi})$ and the covariance matrix has the form 
\begin{equation}
    C = \begin{bmatrix}
\sigma^{2}(\theta) & \sigma_{\times}^{2}(\theta) & 0 \\
\sigma_{\times}^{2}(\theta) & \sigma_{\theta}^{2}(\theta) & 0 \\
0 & 0 & \sigma_{\phi}^{2}(\theta)
\end{bmatrix} \, . 
\end{equation}
Finally we can write the ensemble average of the Minkowski tensor components as
\begin{eqnarray} \langle W_{1}^{0,2}|_{\theta\theta} \rangle &=& {1 \over 2} \int_{\theta} \int \d \delta \d\delta_{\theta} \d \delta_{\phi} P(\delta, \delta_{\theta},\delta_{\phi}) {\delta_{\theta}^{2}  \over \sqrt{\delta_{\theta}^{2} + \delta_{\phi}^{2}} } \delta_{D}(\delta - \nu)  \sin\theta \d \theta   \, , \\
\langle W_{1}^{0,2}|_{\phi\phi} \rangle &=& {1 \over 2} \int_{\theta} \int \d \delta \d\delta_{\theta} \d \delta_{\phi} P(\delta, \delta_{\theta},\delta_{\phi}) {\delta_{\phi}^{2}  \over \sqrt{\delta_{\theta}^{2} + \delta_{\phi}^{2}} } \delta_{D}(\delta - \nu)  \sin\theta \d \theta \, , \\ 
\langle W_{1}^{0,2}|_{\theta\phi}\rangle  &=& 0 \, ,
\end{eqnarray}
where the off-diagonal element is zero due to the $\delta_{\phi}$ integrand being an odd function distributed about zero, independently of $\delta_{\theta}$ and we have used $\int_{\theta}$ to denote an angular integration over some range of $\theta$ (for the whole sphere $\int_{\theta} = \int_{0}^{\pi}$). Also of interest in this section is the trace of the tensor, which is simply the scalar Minkowski functional $W_{1}$ : 
\begin{equation} \langle W_{1} \rangle = {1 \over 2} \int_{\theta} \int \d \delta \d\delta_{\theta} \d \delta_{\phi} P(\delta, \delta_{\theta},\delta_{\phi})  \sqrt{\delta_{\theta}^{2} + \delta_{\phi}^{2}}  \delta_{D}(\delta - \nu)  \sin\theta \d \theta \, .
\end{equation}
The integrals over the random variables $\delta, \, \delta_{\theta}, \, \delta_{\phi}$ run over $(-\infty,\infty)$. The $\delta$ integral is trivial due to the delta function, and the joint Gaussian PDF reduces to a separable, conditional probability $P(\delta_{\theta}, \delta_{\phi}|\delta = \nu) = P(\delta_{\theta}|\delta = \nu) P(\delta_{\phi})$, where $P(\delta_{\theta}|\delta = \nu) = {\cal N}\left({\sigma_{\times}^{2} \over \sigma^{2}}\nu, \sigma_{\theta}^{2} - (\sigma_{\times}^{2})^{2}/\sigma^{2}\right)$ and $P(\delta_{\phi}) = {\cal N}(0,\sigma_{\phi}^{2})$. The conditional probability distribution can be written as $P(\delta_{\theta}|\delta = \nu) = {\cal N}\left({\sigma_{\times}^{2} \over \sigma^{2}}\nu, \sigma_{\phi}^{2}\right)$, meaning that the distributions describing $\delta_{\theta}$ and $\delta_{\phi}$ differ only by the shifted mean generated by the correlation $\langle \delta \delta_{\theta}\rangle$. The ensemble averages can be written as 
\begin{eqnarray} \langle W_{1}^{0,2}|_{\theta\theta} \rangle &=& {1 \over 2} \int_{\theta} {e^{-\nu^{2}/2\sigma^{2}} \over \sqrt{2\pi}\sigma} \int_{-\infty}^{\infty} \int_{-\infty}^{\infty} \d\delta_{\theta} \d \delta_{\phi} P(\delta_{\theta}|\nu) P(\delta_{\phi})  {\delta_{\theta}^{2}  \over \sqrt{\delta_{\theta}^{2} + \delta_{\phi}^{2}} }   \sin\theta \d \theta  \, ,  \\
\langle W_{1}^{0,2}|_{\phi\phi} \rangle &=& {1 \over 2} \int_{\theta} {e^{-\nu^{2}/2\sigma^{2}} \over \sqrt{2\pi}\sigma} \int_{-\infty}^{\infty} \int_{-\infty}^{\infty} \d\delta_{\theta} \d \delta_{\phi} P(\delta_{\theta}|\nu) P(\delta_{\phi})  {\delta_{\phi}^{2}  \over \sqrt{\delta_{\theta}^{2} + \delta_{\phi}^{2}} }   \sin\theta \d \theta  \, ,  \\ 
\langle \Delta W_{1}^{0,2}\rangle &\equiv& \langle W_{1}^{0,2}|_{\theta\theta} \rangle - \langle W_{1}^{0,2}|_{\phi\phi} \rangle = {1 \over 2} \int_{\theta} {e^{-\nu^{2}/2\sigma^{2}} \over \sqrt{2\pi}\sigma} \int_{-\infty}^{\infty} \int_{-\infty}^{\infty} \d\delta_{\theta} \d \delta_{\phi} P(\delta_{\theta}|\nu) P(\delta_{\phi})  {\delta_{\theta}^{2} -  \delta_{\phi}^{2} \over \sqrt{\delta_{\theta}^{2} + \delta_{\phi}^{2}} }   \sin\theta \d \theta  \, , \\
\label{eq:w1} \langle W_{1} \rangle &=& \langle W_{1}^{0,2}|_{\theta\theta} \rangle + \langle W_{1}^{0,2}|_{\phi\phi} \rangle = {1 \over 2} \int_{\theta} {e^{-\nu^{2}/2\sigma^{2}} \over \sqrt{2\pi}\sigma} \int_{-\infty}^{\infty} \int_{-\infty}^{\infty} \d\delta_{\theta} \d \delta_{\phi} P(\delta_{\theta}|\nu) P(\delta_{\phi})  {\delta_{\theta}^{2} +  \delta_{\phi}^{2} \over \sqrt{\delta_{\theta}^{2} + \delta_{\phi}^{2}} }   \sin\theta \d \theta  \, .
\end{eqnarray}
The numerator of $\langle\Delta W_{1}^{0,2}\rangle$ includes the  squared difference of two random variables with the same variance, but different means. It will be negligible if the condition $\lambda/(\sigma_{1}/\sigma_{0}) \ll 1$ holds. This is the case for the example considered in the main body of this work, which has $\lambda/(\sigma_{1}/\sigma_{0}) \simeq 0.01$. 

For perturbative dipole modulation $\lambda \ll 1$, we can calculate the leading order effect on $\langle W_{1}\rangle$ and $\langle \Delta W_{1}^{0,2} \rangle$ exactly. Expanding all terms in $\lambda \ll 1$, we find the following leading order behaviour 
\begin{eqnarray} \label{eq:dip_dw} \langle \Delta W_{1}^{0,2} \rangle &=& {9 \over 128\pi} \lambda^{2} {\sqrt{2}\sigma_{0} \over \sigma_{1}} \nu^{2} e^{-\nu^{2}/2} \int_{\theta} \sin^{3}\theta d\theta  + {\cal O}(\lambda^{3}) \, , \\
\label{eq:dip_w1} \langle W_{1} \rangle &=& {\sigma_{1} \over 4 \sqrt{2}\sigma_{0}} e^{-\nu^{2}/2} \int_{\theta} \left( 1 + \lambda   \sqrt{3 \over 4\pi} \nu^{2} \cos\theta\right) \sin\theta d\theta + {\cal O}(\lambda^{2}) \, ,
\end{eqnarray}
where we have not integrated out the explicit $\theta$ dependence. We see that the traceless part $\langle \Delta W_{1}^{0,2} \rangle$, which is the additional information that the Minkowski tensors are sensitive to compared to the scalar functionals, is second order in the anisotropy parameter $\lambda$. The trace of the tensor, which is the scalar functional $\langle W_{1} \rangle$ in equation (\ref{eq:dip_w1}), picks up an additional contribution that is linear in $\lambda$, which generates a variance modulation characterised by the $\nu^{2}$ threshold dependence. This term is present despite the field being Gaussian and is a direct consequence of non-stationarity. However, this leading order contribution integrates to zero over $S^{2}$ due to the angular integral in (\ref{eq:dip_w1}) when performed over $\theta \in [0,\pi]$. If we measure the Minkowski functional $W_{1}$ averaged over $S^{2}$ then the dipole modulation enters at second order in the anisotropy parameter. If we were to instead measure the functionals in fixed $\theta$ bands over the sky, relative to the dipole direction, then we should observe a spatially varying variance beyond the standard global average. In this work the dipole direction is known, in actual data one would have to measure the Minkowski functionals in randomly oriented bands on the sky, and search for systematic changes in the variance beyond random noise. 

\begin{figure}
    \centering
    \includegraphics[width=0.98\textwidth]{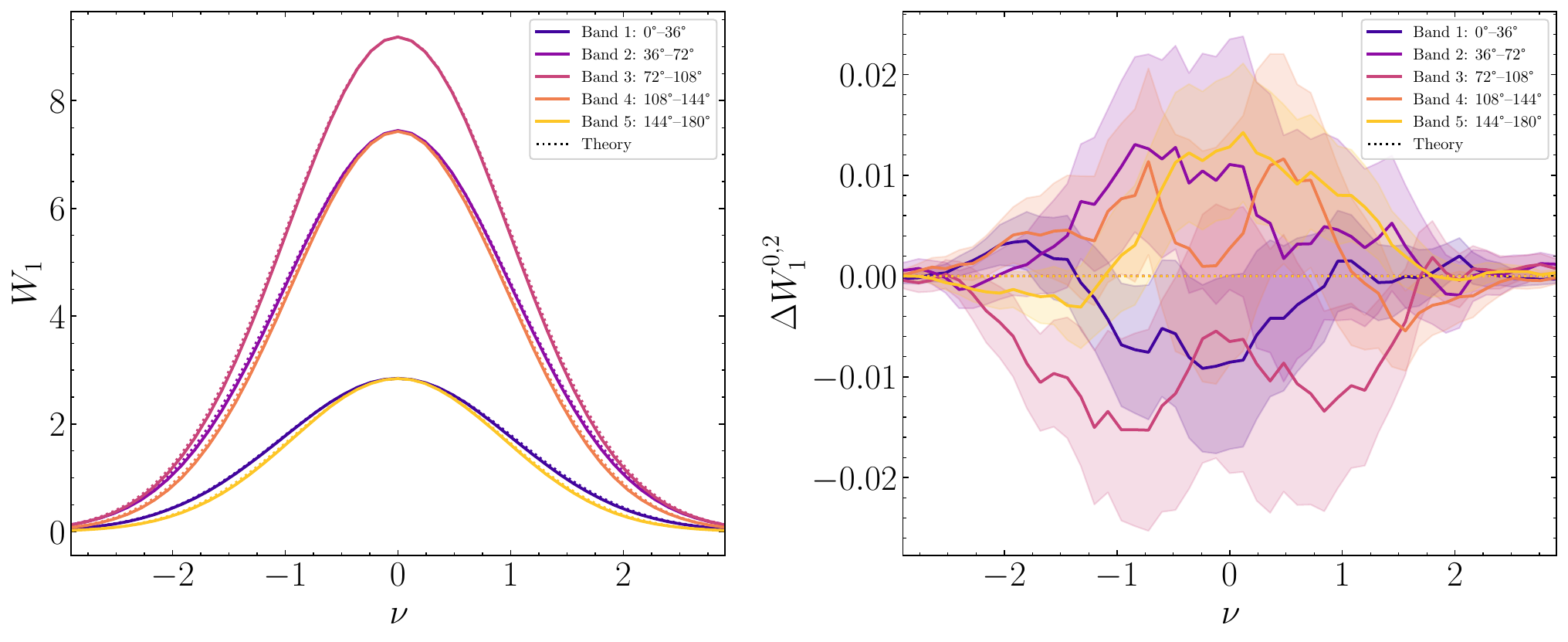} 
\caption{[Left Panel] Minkowski functional $W_{1}$ in five latitude bands extracted from a dipole modulated Gaussian random field with $\lambda = 0.1$. The north/south power asymmetry generates a band variance modulation. The same colored dashed lines are the analytic predictions from Section \ref{sec:dip}. [Right Panel] Difference in diagonal components of the Minkowski Tensor, measured in the same five latitude bands as the left panel. There is no statistically significant signal in $\Delta W_{1}^{0,2}$ arising from the dipole modulation, all measurements are consistent with zero. The filled bands correspond to the error on the mean of $N=50$ realisations.}
    \label{fig:W1}
\end{figure}

In the left panel of Figure \ref{fig:W1} we present an example of the signal generated by the dipole modulation in the Minkowski Functional $W_{1}$. We generate $N=50$ random fields with a dipole modulation of amplitude $\lambda = 0.1$, and measure $W_{1}$ in five bands relative to the dipole direction (north-south), with `band 3' running along the equator. There are two pairs of curves : bands 1/5 (dark blue/yellow), and bands 2/4 (purple/orange), which are of equal area, and located in the north/south sky respectively. These $W_{1}$ paired measurements (solid lines) are characterised by equal amplitude but differing effective variance, which is due to the effect of the $\lambda$ term in equation (\ref{eq:dip_w1}). Band 3 covers the equator and has no matching curve. The dashed lines correspond to the theoretical expectation value calculated in (\ref{eq:dip_w1}), where the angular integral has been performed over the bands. We find close agreement between the theoretical expectation and numerical reconstruction (solid lines and the same coloured dashed lines). The errors on the means of the numerical measurements are plotted as coloured bands, but are not discernible by eye. In the right panel we present $\Delta W_{1}^{0,2}$ in the same bands as the left panel, with the dashed lines representing the expectation value in equation (\ref{eq:dip_dw}). The solid colored bands represent the error on the mean, which are practically consistent with zero. There is no detectable signal in $\Delta W_{1}^{0,2}$, as expected from the second order nature of the expectation value. The anisotropic signal is simultaneously suppressed by the factors $9/128\pi \ll 1$, $\lambda^{2} \ll 1$ and $\sigma_{0}/\sigma_{1} \ll 1$.

In conclusion, dipole modulation does generate a signal in the Minkowski tensors, but the leading order effect integrates to zero when averaged over the entire sphere. The impact of the modulation on the traceless component of the tensor is a second order effect in the parameter $\lambda$, and is suppressed by $\lambda^{2} \sigma_{0}/\sigma_{1}$ even when $\lambda \sim {\cal O}(1)$. The dominant signal lives in the trace $W_{1}$, which is linear in $\lambda$. This linear signal is a consequence of normalising by the global rather than the band variance. It is equivalent, at leading order, to measuring the one-point variance $\sigma^{2}(\theta)$ in bands, and would vanish under local band normalisation.

More generally, random fields that are modulated by a deterministic, angular dependent function are non-stationary on $S^{2}$, and we should consider carefully which statistic is most sensitive to the signal, and also over which sub-domain of $S^{2}$ we should measure the statistics to maximize the constraining power. For an agnostic search, measuring a multitude of statistics in patches is a sensible approach \cite{Goyal:2021nun,Bashir:2025gen,CarronDuque:2026unb}.

\section{Conclusions}
\label{sec:conclusion}

In this work we have considered how to extract anisotropic signals from random fields defined on $S^{2}$, using the Minkowski tensors. The definition of the MT involves an integral over excursion set boundaries $\delta({\bf x}) = \nu\sigma$. For the Minkowski Functionals, this integration can be performed without any notion of transport because of their scalar nature. However for the tensors, their definition is ambiguous without including some prescription for mapping tangent spaces. This ambiguity is not unique to $S^{2}$, but is a generic feature of defining tensor valued statistics on any manifold. Even in ${\mathbb R}^{2}$ a transport prescription must be specified, but in this case Euclidean translation is trivial and often invoked implicitly. Here, we have highlighted the various problems that arise when attempting to generalise to a curved manifold like $S^{2}$. 

The estimator for $W_{1}^{0,2}$ defined in equation (\ref{eq:def_3}) and discussed in Section \ref{sec:noncov}, is a direct sum of tensor components defined in different tangent spaces, with the local basis ${\bf e}_{\theta}$, ${\bf e}_{\phi}$ implicitly identified across the sphere. The eigenvalue ratio $\alpha = \Lambda_{2}/\Lambda_{1}$ can present a signal in the presence of global anisotropic shear (cf. lower left panel, Figure \ref{fig:alpha}). However, breaking covariance by summing vectors in different tangent spaces compromises the utility of the statistic. Specifically, a local coordinate transformation can spuriously decrease or increase $\alpha$. We stress that this is not merely a cautionary tale of interpretation. In Appendix \ref{app:stereographic} we demonstrate numerically that the non-covariant $\alpha$ can be driven to a false null for a genuinely anisotropic field by an arbitrary choice of local frame. 

We introduced a covariant form of the Minkowski tensor in Section \ref{sec:cov}, by parallel transporting the relevant vectors $\hat{n}$ to a common location on $S^{2}$ prior to averaging. Both the trace and eigenvalues of this estimator are now coordinate invariant. However, we found that anisotropic signals are partially washed out by the geometric rotation generated by the transport. This was confirmed in our numerical results, in particular the isotropic biasing is clearly presented in Figure \ref{fig:alpha}. 

Based on the above discussion, we conclude that neither one-point statistic serves as a satisfactory standalone test of statistical isotropy\footnote{However, if one intends to summarize the data with a single, global one-point statistic, the covariant definition should be preferred because it constitutes a coordinate-independent geometric observable.} on $S^{2}$, motivating the connected-component correlation functions $\xi_{\pm}(\theta,\nu)$ introduced in Section \ref{sec:correlation}. These provide an intrinsic, unbiased test of statistical isotropy on $S^{2}$ that can measure scale-dependent anisotropic signals. We extracted these statistics from four ensembles of fields -- isotropic, dipole modulated, and two shear-types. The first shear  generated a globally preferred direction at each point, pointing away from the north pole, and the second was constructed from a large scale field gradient, which is locally coherent over a few tens of degrees.

Even for statistically isotropic fields, we find non-zero correlations in $\xi_{\pm}$ at small angular separations $\theta \lesssim 5\theta_{G}$, reflecting the local filamentary structure of the fields. At separations $\theta \gtrsim \theta_{G}$, excursion sets are preferentially elongated toward neighbouring connected components, generating positive correlations in both functions that peak at $\theta \simeq 1.5\theta_{G}$. At sub-smoothing separations $\theta < \theta_{G}$ the sign of $\xi_{+}$ reverses while $\xi_{-}$ remains positive, identifying mirror configurations in which the principal axes are tilted at opposite angles to the connecting geodesic. Pairs that remain distinct at these separations are necessarily separated by a saddle point lying well away from the connecting geodesic, toward which both components elongate; the measured tilt increases with threshold, approaching the maximal mirror configuration $\bar{\psi} = 45^{\circ}$ at $\nu = 2$.

For the anisotropic fields, we find a significant excess correlation in $\xi_{+}$ for both shear types. The globally sheared field generates a coherent signal that decays slowly with separation and migrates into $\xi_{-}$. This occurs because the geodesic connecting a pair rotates relative to the alignment pattern as the separation grows, converting direction-coherent alignment into mirror alignment about the connecting arc. The two functions satisfy the mirror relation $\xi_{-}(\theta) \simeq \xi_{+}(180^{\circ}-\theta)$ and cross at $\theta = 90^{\circ}$. In contrast, the locally sheared field generates a $\xi_{+}$ signal that falls by an order of magnitude within $\theta \sim 30^{\circ}$, reflecting the finite coherence of the alignment direction, together with a suppressed $\xi_{-}$ migration sourced by the sub-dominant coherent component that the $\tilde{Q}$ quadratic construction generates. The decay scale of $\xi_{+}$ therefore measures the angular coherence of the alignment, while the antipodal amplitude of $\xi_{-}$ measures its globally coherent fraction. A cross-type anisotropy, rotated $45^{\circ}$ relative to the shear patterns considered here, would generate a $\xi_{-}$ signal of opposite sign; we intend to study this anisotropy type in future work.

%For the anisotropic fields, we find a significant excess correlation in $\xi_{+}$ for both shear types, extending to large angular scales. The globally sheared field generates a coherent signal out to $\theta \gtrsim 30^{\circ}$, with some suppression at the largest separations due to the geometric rotation of ${\bf e}_{\theta}$ as one moves around the sphere. Specifically, the locally defined north-south direction de-correlates at large angular distances. In contrast, the locally sheared field generates a qualitatively similar but distinct signal in $\xi_{+}$, which decays to zero beyond the coherence scale of the modulating field $\theta_{\rm mod} \sim 36^{\circ}$. This highlights the scale-dependent discriminating power of these statistics. Neither shear type generates a significant signal in $\xi_{-}$ beyond the smallest angular bin. A positive $\xi_{-}$ signal would be generated for tangential alignments relative to fixed points and we intend to study this anisotropy type in future work. 

Perhaps the most pertinent anisotropic signal in modern cosmology is dipole modulation, however we found that this case generates no statistically significant signal in $\alpha$ nor the correlation functions $\xi_{\pm}$. We examined this case analytically, and found that the impact of the dipole modulation on the traceless component of $W_{1}^{0,2}$ is suppressed by a factor of $\lambda^{2} \sigma_{0}/\sigma_{1} \ll 1$ (cf. equation (\ref{eq:dip_dw})). However, the trace of $W_{1}^{0,2}$ does pick up a signal that is linear in $\lambda$, with a distinct variance modulation generated in ${\rm Tr}\left(W_{1}^{0,2} \right) = W_{1}(\nu)$ when measured in bands on the sky. This result highlights that there is no single statistic that should be adopted to measure an anisotropic signal -- $\Delta W_{1}^{0,2}$, $W_{1}$, $\xi_{\pm}(\theta,\nu)$ and other quantities can  be extracted from the data, and each provides some information on the nature of the field.

Searching for anisotropic signals in random fields defined on $S^{2}$ is a delicate business. Fields living in ${\mathbb R}^{D}$ can be statistically anisotropic but homogeneous, and hence stationary. An example is the plane-parallel, redshift space distorted matter density in ${\mathbb R}^{3}$, which is characterised by spatially constant cumulants \cite{1996ApJ...457...13M}\cite{Codis:2013exa}\cite{Appleby:2025sbk}. For fields living on $S^{2}$, breaking the underlying ${\rm SO}(3)$ symmetry can generate non-stationary cumulants, meaning that the statistical properties of the field will become position-dependent. Since we necessarily relate spatial and ensemble averages in cosmology, the domain over which we should spatially average depends on the signal that we are searching for. We have presented a simple example in this work -- we showed how a dipole modulated field generates a second-order signal if we measure an all-sky average of the Minkowski tensors, but a linear signal can be extracted if we average the trace in latitude bands relative to the dipole direction. Such a scenario is specific to the dipole, and one can expect different types of anisotropy will require specific averaging patterns. The look-elsewhere effect should be carefully considered in this context, because the domain of spatial averaging is a free `parameter' that can be tuned to find a signal. 

In this work we have considered two distinct types of anisotropy. Dipole modulation corresponds to a non-stationary field, with position-dependent cumulants. The shear fields are stationary, but isotropy is broken by introducing a preferred direction in the field derivatives, which are coupled to a spin-2 field. Different types of anisotropy can be related to different subgroup breakings of the underlying $SO(3)$ symmetry, in exact analogy with the breaking of translation and rotation invariance on ${\mathbb R^{2}}$. A full accounting of the different symmetry breaking mechanisms, and what signal the breaking will generate in our statistics, will be considered in future work. Also deferred to the future is the application of our methodology to data -- the Cosmic Microwave Background and two-dimensional projections of the galaxy distribution.

\acknowledgments{SA is supported by a KIAS Individual Grant (PG055703) at Korea Institute for Advanced Study (KIAS). C.B.P. is supported by KIAS individual grant PG016903 and by the National Research Foundation of Korea (NRF) grant funded by the Korean government (MSIT; RS-2024-00360385). SA would like to thank Atsushi Taruya, Christophe Pichon, Pravabati Chingangbam and Priya Goyal for helpful discussions during the writing of this manuscript. Codes used in this work will be made available after publication. }

\appendix

\section{Numerical Methodology} 
\label{app:numerics_I}

We start by placing $N_{\rm p}$ points, randomly and isotropically distributed, on the sphere 
\begin{equation} \phi_{i} = 2\pi u_{i} \qquad \theta_{i} = \cos^{-1} (2v_{i} - 1)  \, ,
\end{equation} 
where $u_{i}, v_{i}$ are uniformly distributed variables $\sim {\cal U}(0,1]$. The $i^{\rm th}$ point has corresponding unit vector ${\bf p}_{i} = (\sin\theta_{i}\cos\phi_{i}, \sin\theta_{i}\sin\phi_{i},\cos\theta_{i})$ in $\mathbb{R}^{3}$. Next, we generate a spherical Delaunay triangulation of the points using a convex hull algorithm. 

Next, we dress the points ${\bf p}_{i}$ with density values drawn from a Gaussian random distribution. Specifically, we define a power spectrum $C_{\ell} = \ell^{n} \exp[-\ell (\ell+1) \theta_{G}^{2}]$, where $\theta_{G}$ is some smoothing scale and $n$ is input. We define $a_{\ell m} = \sqrt{C_{\ell}} x$, $b_{\ell m} = \sqrt{C_{\ell}} y$, where $x,y$ are unit variance Gaussian distributed random variables, and define the density field at ${\bf p}_{i}$ as 

\begin{equation}
\delta_{i} = \sum_{\ell=1}^{\ell_{\rm max}} \left[ a_{\ell 0} A_{\ell 0} P_{\ell 0}(\cos\theta_i) + \sum_{m=1}^{\ell} \sqrt{2} A_{\ell m} P_{\ell
  m}(\cos\theta_i) \left[ a_{\ell m} \cos (m\phi_{i}) + b_{\ell m} \sin (m\phi_{i}) \right] \right]
  \end{equation}

where $P_{\ell m}(\cos\theta_{i})$ are the associated Legendre polynomials and $A_{\ell m}$ are the spherical harmonic normalisation coefficients 
\begin{equation}
 \label{eq:norm}   A_{\ell m} = \sqrt{{(2\ell + 1) \over 4\pi} {(\ell - m)! \over (\ell + m)!}} \, ,
\end{equation}
defined such that the real spherical harmonics form an orthonormal basis on $S^{2}$. We unit-normalise this field $\delta_{i} \to \delta_{i}/\bar{\sigma}$. 

We define a set of $N_{\rm t} = 51$ threshold values equi-spaced over the range $-3 \leq \nu \leq 3$. For a given threshold, we find all points for which $\delta_{i} > \nu$ : these approximately form the excursion set. In what follows we call points at which $\delta_{i} > \nu$ as `in' and $\delta_{i} < \nu$ as `out' of the set. The Minkowski Functionals, tensors and Betti numbers are then estimated using the method of marching triangles\footnote{See \cite{Collischon:2024jhw} for an alternative numerical prescription.}. Specifically, each triangle in the spherical Delaunay construction has three corresponding $\delta_{k}, \delta_{k+1}, \delta_{k+2}$ density values at its vertices ${\bf p}_{k}, {\bf p}_{k+1}, {\bf p}_{k+2}$. The statistics are constructed algorithmically according to the following rules :  

\begin{enumerate} 
\item  If $\delta_{k}, \delta_{k+1}, \delta_{k+2} > \nu$, then the triangle area is added to $W_{0}$. 

\item If any two of the density values satisfy $\delta_{k}, \delta_{k+1} > \nu$ and $\delta_{k+2} < \nu$, then we linearly interpolate, on the surface of $S^{2}$, using great arc distances, along the triangle edges $[\delta_{k}, \delta_{k+2}]$ and $[\delta_{k+1},\delta_{k+2}]$ to find the two points ${\bf t}_{1}$, ${\bf t}_{2}$ at which $\delta = \nu$. The area enclosed by the four vertices $({\bf t}_{1}, {\bf t}_{2}, {\bf p}_{k}, {\bf p}_{k+1})$ is added to $W_{0}$ and the length of the great arc segment ${\bf e}_{12}$ between $[{\bf t}_{1},{\bf t}_{2}]$ on $S^{2}$ is added to $W_{1}$. The points $[{\bf t}_{1}, {\bf t}_{2}]$ form a great arc, at the endpoint ${\bf t}_{1}$ (chosen arbitrarily) we construct the unit normal ${\bf \hat{n}}$ to the great arc direction in the tangent plane of $S^{2}$ at this point. The Minkowski Tensor components are then updated as $W_{1}^{0,2} \to W_{1}^{0,2} + |{\bf e}_{12}| ({\bf \hat{n}} \otimes {\bf \hat{n}})$. Note that the normal direction ${\bf \hat{n}}$ will vary depending on where along the great arc segment $[{\bf t}_{1}, {\bf t}_{2}]$ we evaluate it. We neglect this variation on the grounds that $|{\bf e}_{12}| \ll 1$.

\item If one density value satisfies $\delta_{k} > \nu$ and the other two $\delta_{k+1}, \delta_{k+2} < \nu$, then we linearly interpolate along the triangle edges $[\delta_{k}, \delta_{k+1}]$ and $[\delta_{k},\delta_{k+2}]$ to find the two points ${\bf t}_{1}$, ${\bf t}_{2}$ at which $\delta = \nu$. The area enclosed by the three vertices $({\bf t}_{1}, {\bf t}_{2}, {\bf p}_{k})$ is added to $W_{0}$ and the length $|{\bf e}_{12}|$ of the segment $[{\bf t}_{1},{\bf t}_{2}]$ on $S^{2}$ is added to $W_{1}$. Similarly to the previous case, the normal ${\bf n}$ to the great arc segment $[{\bf t}_{1}, {\bf t}_{2}]$ is evaluated at ${\bf t}_{1}$ and the Minkowski Tensor components are updated as $W_{1}^{0,2} \to W_{1}^{0,2} + |{\bf e}_{12}| ({\bf \hat{n}} \otimes {\bf \hat{n}})$. 

\item If all density values satisfy $\delta_{k}, \delta_{k+1}, \delta_{k+2} < \nu$, then the triangle does not contribute to the MF, MTs. 
\end{enumerate}
The Betti numbers are calculated using a graph search on adjacent triangles. Each point ${\bf p}_{i}$ in the excursion set is initially assigned a label $n=0$ to mark it as unvisited in the search. We then iterate through all triangles sequentially. When a triangle is encountered that contains at least one `in' vertex with $n=0$, a connected component integer $n_{c}$ is incremented $n_{c} \to n_{c}+1$ and all `in' vertices of this triangle are assigned the label $n=n_{c}$. The search then propagates to all neighbouring triangles sharing an edge, similarly assigning $n=n_{c}$ to any $n=0$ indices. This procedure is repeated iteratively on triangles sharing edges until no more $n=0$ are found. The sequential scan through triangles is then resumed, and the process is repeated once the next $n=0$ labeled point is found. Once all `in' vertices have been assigned a non-zero label, the total number of distinct components $n_{c}$ is equal to the Betti number $b_{0}$ at the threshold $\nu$. The identical algorithm applied to `out' vertices $\delta_{k} < \nu$ yields $b_{1}$. The Minkowski functional $W_{2}$, the genus, is then $W_{2} = (b_{0} - b_{1})/4\pi$.

For the purpose of our study, for each threshold $\nu$ every vertex in the excursion set has now been assigned an integer label $1 \leq n \leq b_{0}$. We can therefore use this assignment to construct the Minkowski Functionals and tensors for every connected component within the excursion set. For each connected component $\mu$, we define its `geometric center' as the mean ${\bf \bar{p}}_{\mu} = \sum_{\eta \in \mu} {\bf p}_{\eta}$, normalised to a unit vector. We then define $W_{1 , \mu}^{0,2} = \sum_{{\rm edges} \in \mu} |{\bf e}_{12}| ({\bf \hat{n}} \otimes {\bf \hat{n}})$ as the Minkowski tensor $W_{1}^{0,2}$ for each $\mu$ connected component, which we evaluate at ${\bf \bar{p}}_{\mu}$. There is a geometric rotation present here, in the sense that we transport ${\bf \hat{n}}$ to ${\bf \bar{p}}_{\mu}$ prior to taking the average of $({\bf \hat{n}} \otimes {\bf \hat{n}})$. If the excursion sets are suitably small relative to the curvature of $S^{2}$, then this should not be a significant factor. However, we consider the magnitude of this effect in the following subsection. Neglecting this complication for now, we diagonalize each $W_{1}^{0,2}|_{\mu}$, defining the eigenvalues and corresponding vectors as $\Lambda_{1, \mu} \geq  \Lambda_{2, \mu}$ and ${\bf v}_{\mu}$, ${\bf u}_{\mu}$. The dimensionless shape parameter $\beta_{\mu} = \Lambda_{2, \mu}/\Lambda_{1, \mu} \leq 1$ describes the ellipticity of the $\mu^{\rm th}$ connected component.

\section{finite-size geometric distortion}

The finite-size geometric distortion is the warping of the connected component shapes due to the action of parallel transporting the perimeter normals to the geometric centers ${\bf \bar{p}}_{\mu}$, prior to extracting the eigenvalues of the Minkowski tensor and inferring $\beta_{\mu}$. We expect the pullback to a common point to slightly isotropize the objects for the same reason that the covariant definition of the global Minkowski tensor is isotropized by geodesic transport, and in this appendix we quantify this effect. To do so, we construct the Minkowski tensors of the connected components at ${\bf \bar{p}}_{\mu}$ using two methods -- our standard great arc transport and non-covariant averaging of the components without transport -- and comparing the shape parameter $\beta_{\mu}$ as a function of size of connected component. We repeat this test for multiple thresholds $\nu$. We define the fractional difference 
\begin{equation} 
\Delta \beta_{\mu} = {\beta_{\mu}^{({\rm cov})} - \beta_{\mu}^{({\rm non-cov})} \over \beta_{\mu}^{({\rm cov})}} \, ,
\end{equation}
where $\beta_{\mu}^{({\rm cov})}$ and $\beta_{\mu}^{({\rm non-cov})}$ are the connected component shape parameters constructed using geodesic great arc transport and summing the components in different tangent spaces respectively. 

\begin{figure}
    \centering
    \includegraphics[width=0.98\textwidth]{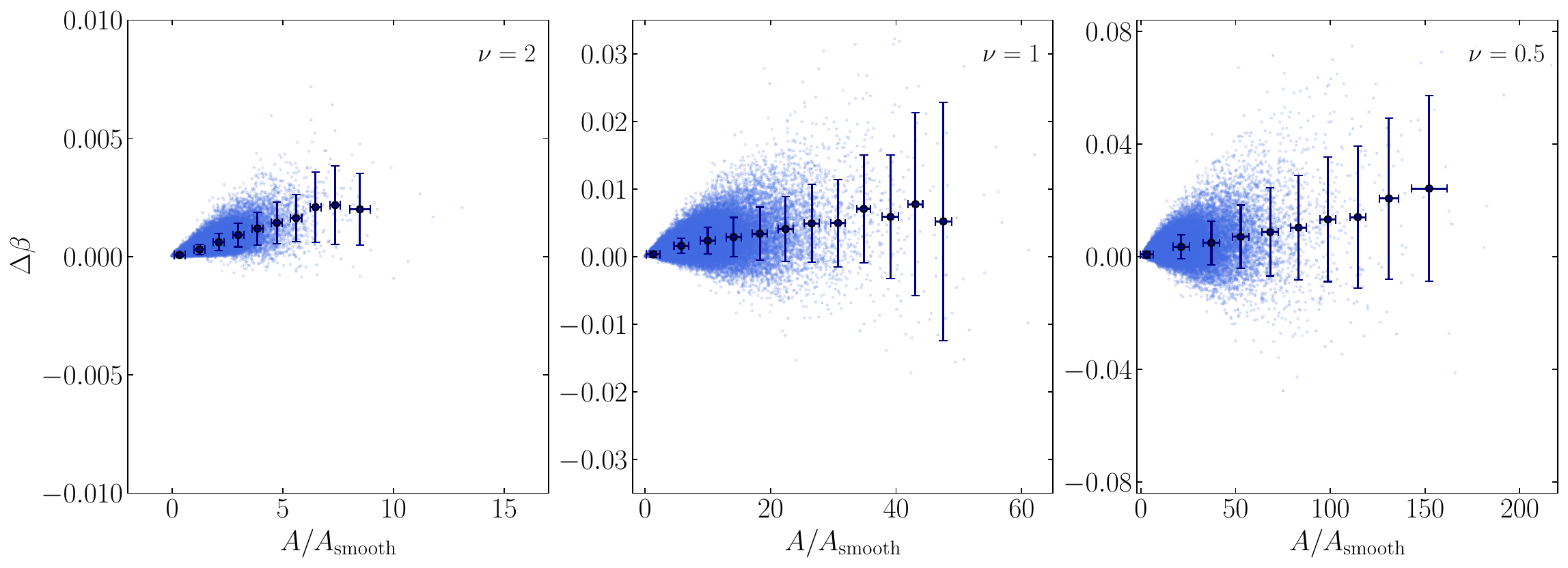} 
\caption{The fractional difference in connected component shape parameter $\Delta \beta$ (covariant vs non-covariant formulation) as a function of size of the connected component (normalised by the smoothing area $A_{\rm smooth} = \pi \theta_{G}^{2}$). The left/middle/right panels are connected components measured at $\nu=2$, $\nu=1$ and $\nu = 0.5$ respectively. Each small blue point is a single connected component, and the large dark blue points/error bars are the binned average of the point clouds.}
    \label{fig:pb}
\end{figure}

In Figure \ref{fig:pb} we present $\Delta \beta_{\mu}$ as a function of $A_{\mu}/A_{\rm smooth}$ for every connected component extracted from $N=10$ realisations of an isotropic random field on $S^{2}$. We have defined the smoothing area as $A_{\rm smooth} = \pi \theta_{G}^{2}$, where $\theta_{G}$ is the Gaussian smoothing scale. Each small blue point is a single connected component, and the large navy points/error bars are the binned average of the point clouds. The left/middle/right panels are connected components measured at density thresholds $\nu=2$, $\nu=1$, $\nu=0.5$ respectively. 

In each of the panels, we find $\Delta \beta$ is systematically biased towards positive values, and increases with increasing connected component size. This is expected : we have already seen that the act of transport isotropizes the global Minkowski tensors. However, the effect is small because the transport is short compared to the curvature for most connected components. For $\nu =2$, the size of the systematic distortion is $< 0.5\%$ and is less than $< 1\%$ for most objects for $\nu=1$ and $\nu=0.5$. Only for the largest objects, at low thresholds, does the shape change become greater than $1\%$. The correlation functions $\xi_{\pm}$ are pair counts, and the size of the connected components are irrelevant for the construction of the statistic. It follows that the systematic shift in $\beta_{\mu}$ for the large connected components will not strongly affect the statistics, which will be dominated by the more numerous, smaller objects.

\section{Stereographic Projection}
\label{app:stereographic}

Early works on the Minkowski Functionals extracted the statistics from fields on $S^{2}$ by first projecting the field onto the plane using a conformal mapping \citep{Gott:1989yj}. This methodology was subsequently adopted in more recent work \citep{Ganesan:2017} for the tensors. In this appendix we briefly discuss the application of the stereographic projection to the tensors.

We define the stereographic projection from the south pole $(0,0,-1)$ onto the plane tangent to the north pole, using the transformation of a point $(\sin\theta \cos\phi, \sin\theta\sin\phi,\cos\theta)$ to $\rho = 2\tan(\theta/2)$, $\phi' = \phi$, where $\rho$ and $\phi'$ are the radial and azimuthal coordinates on the plane. The point $\theta=\pi$ is a coordinate singularity of the mapping, but is measure zero and not important for the following discussion. Because the azimuthal angles are transformed to each other one-to-one, and the mapping is conformal (angle preserving), the relative weighting of the $\theta$ and $\phi$ components is preserved and a unit vector $\hat{n} = n_{\theta} {\bf e}_{\theta} + n_{\phi} {\bf e}_{\phi}$ on $S^{2}$ is mapped exactly to $\hat{n}' = n_{\theta} {\bf e}_{\rho} + n_{\phi} {\bf e}_{\phi'}$ in the plane. There is an overall length shift of the vector by a factor of $\Lambda(\theta) = {\rm sec}^{2}(\theta/2)$, but this is absorbed into the normalisation of the basis vectors. In the definition of the Minkowski tensor $W_{1}^{0,2}$, there is an additional scaling of the perimeter length. Specifically, the perimeter integral $\int d\ell$ on $S^{2}$ maps to an elongated boundary $\int \Lambda({\theta}) d\ell$ on $R^{2}$, and so excursion set edges must be re-scaled by $1/\Lambda(\theta)$ on the plane to infer the Minkowski functionals on the sphere. 

\begin{figure}
    \centering
    \includegraphics[width=0.98\textwidth]{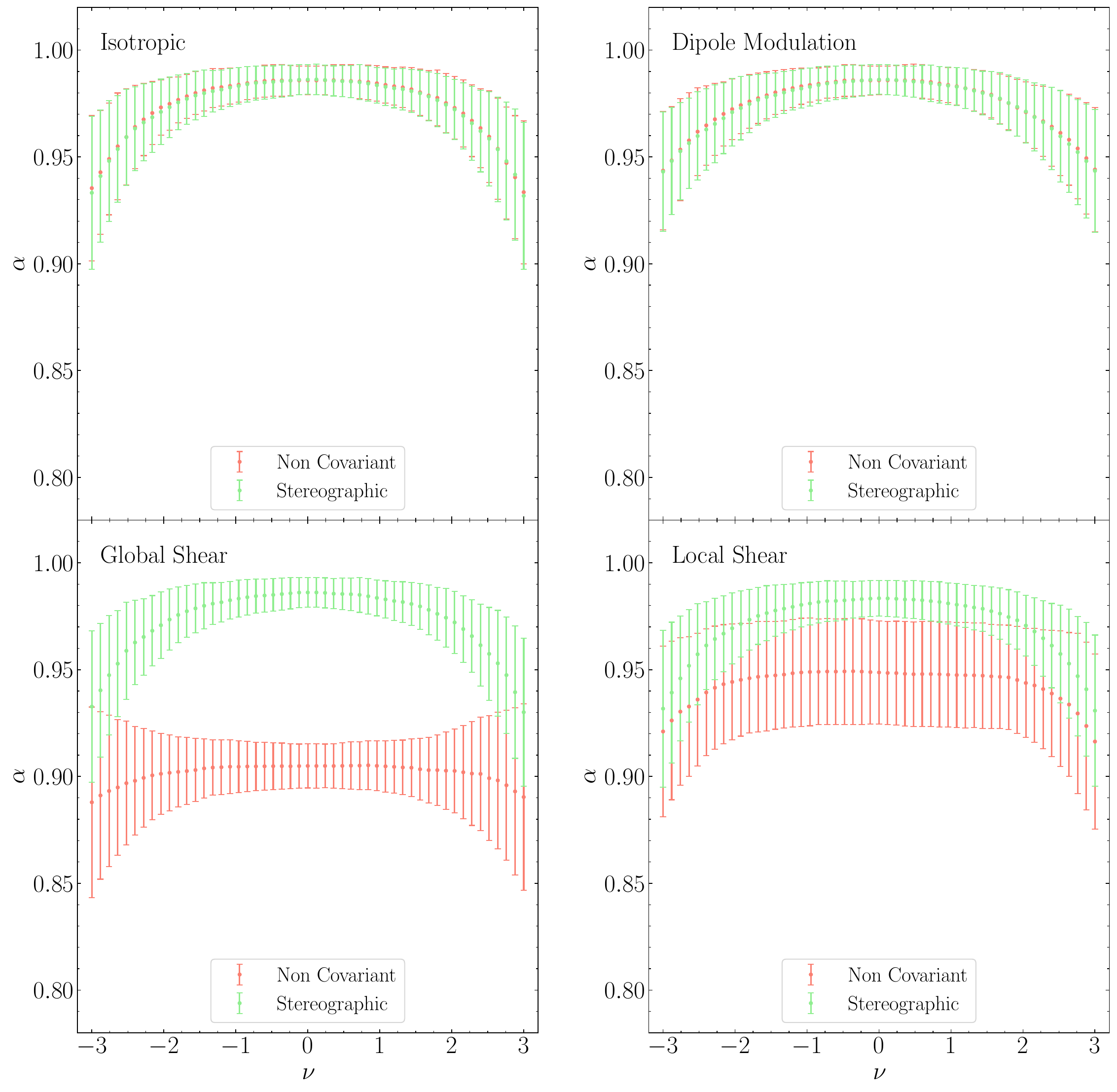} 
\caption{The statistic $\alpha$ extracted from $N_{\rm real}=400$ realisations of the four field ensembles (cf. Figure \ref{fig:alpha}), now comparing the non-covariant pixel sum (red) and stereographic projection measured in Cartesian coordinates on the plane (green). For the globally sheared field (lower left), the stereographic/Cartesian estimator is statistically indistinguishable from the isotropic result. This demonstrates that the non-covariant estimator's detection of a signal is frame-dependent. }
    \label{fig:stereo}
\end{figure}

To calculate $W_{1}^{0,2}$, the act of projecting the field from $S^{2}$ onto the plane, calculating the excursion set boundary normals in the ${\bf e}_{\rho}$, ${\bf e}_{\phi'}$ basis and summing over all pixels is exactly equivalent to the $S^{2}$ pixel sum estimator in Section \ref{sec:noncov}  (cf. equation \ref{eq:sph_0}). Hence, the method inherits the same non-covariant issue as the naive pixel sum on $S^{2}$. Note that once on the plane, one can perform an azimuthal rotation to obtain the unit vector in standard Cartesian coordinates $(n_{x}, n_{y}) = R(\phi')(n_{\rho},n_{\phi'})$, with 
\begin{equation}
  R(\phi') = \begin{pmatrix}
\cos\phi' & -\sin\phi' \\
\sin\phi' & \cos\phi'
\end{pmatrix} \, ,
\end{equation}
however the pixel sum in this basis is still not related to any covariant tensor on $S^{2}$. In fact, if we take the global shear example from the main body of the paper, project the field and measure $\alpha$ in the plane using the ${\bf e}_{\rho}$, ${\bf e}_{\phi'}$ basis, we obtain exactly the same result as the non-covariant estimator in the paper (cf. red points, bottom left panel of Figure \ref{fig:alpha}). However, if we apply the same stereographic projection, and measure the $\alpha$ statistic in the ${\bf e}_{x}$, ${\bf e}_{y}$ coordinate system on the plane, the local azimuthal rotation matrix entirely washes out the anisotropic signal in $\alpha$. This is due to the fact that the rotated projection operator $R(\phi') \, Q \, R(\phi')^{\rm T}$  (where $Q$ is defined in equation \ref{eq:glob_proj}) integrates to zero over a complete azimuthal ring, since $Q$ transforms as a spin-2 object under rotations. We confirm this numerically in Figure \ref{fig:stereo}, where we compute $\alpha$ for all four field ensembles introduced in Section \ref{sec:iso} using the pixel sum on $S^{2}$ (cf. equation (\ref{eq:sph_0}), red points/error bars) and the stereographic projection onto the plane, in the Cartesian ${\bf e}_{x}$, ${\bf e}_{y}$ basis (green points/error bars). For the isotropic and dipole modulated fields, the two estimators agree to within statistical uncertainty. However, for the global shear case (bottom left panel), the stereographic/Cartesian measurement is consistent with isotropy, showing no detectable signal. The two measurements are extracted from the same fields; the only difference is the choice of local coordinate basis and they yield different conclusions. This undermines the utility of $\alpha$ as extracted from the non-covariant estimator.

In spite of this, the signal in the global shear case can in principle be recovered even in the projected frame, by measuring each excursion region's orientation relative to its meridian rather than the fixed Cartesian axis. However, this requires knowledge of the anisotropy axis and is not a blind statistic. This highlights that no single prescription is universally viable and recovering a signal depends on whether the measurement basis happens to match the symmetry of the anisotropy within the field. 

%We have checked numerically that the measurement of $\alpha$ from the projected data and in the ${\bf e}_{x}$, ${\bf e}_{y}$ frame, contains no anisotropic signal, and is confirmation of the non-covariance of the naive pixel sum. 

Note that the trace is still a scalar, coordinate invariant and equal to $W_{1}$. Hence, the stereographic projection method yields exactly the Minkowski Functional $W_{1}$, subject only to lengths being re-scaled by the function $\Lambda(\theta)$.

This discussion highlights the central issue presented in Section \ref{sec:noncov}. The trace of a tensor is insensitive to any pointwise rotation applied before summation, but the tensor itself is not. Therefore any estimator built by summing components in a non-parallel-transported frame correctly recovers the scalar Minkowski functionals but will not yield a covariant tensor. This is the case whether we consider a pixel sum in a spherical basis on $S^{2}$ or a conformal projection onto the plane.

\bibliography{refs}{}

@ARTICLE{2018ApJ...863..200A,
       author = {{Appleby}, Stephen and {Chingangbam}, Pravabati and {Park}, Changbom and {Yogendran}, K.~P. and {Joby}, P.~K.},
        title = "{Minkowski Tensors in Three Dimensions: Probing the Anisotropy Generated by Redshift Space Distortion}",
      journal = {\apj},
         year = 2018,
        month = aug,
       volume = {863},
       number = {2},
          eid = {200},
        pages = {200},
          doi = {10.3847/1538-4357/aacf8c},
archivePrefix = {arXiv},
       eprint = {1805.08752},
 primaryClass = {astro-ph.CO}
}

@article{Planck:2019evm,
    author = "Akrami, Y. and others",
    collaboration = "Planck",
    title = "{Planck 2018 results. VII. Isotropy and Statistics of the CMB}",
    eprint = "1906.02552",
    archivePrefix = "arXiv",
    primaryClass = "astro-ph.CO",
    doi = "10.1051/0004-6361/201935201",
    journal = "Astron. Astrophys.",
    volume = "641",
    pages = "A7",
    year = "2020"
}

@ARTICLE{1996ApJ...457...13M,
   author = {{Matsubara}, T.},
    title = "{Statistics of Isodensity Contours in Redshift Space}",
  journal        = "ApJ.",
     year = 1996,
    month = jan,
   volume = 457,
    pages = {13},
      doi = {10.1086/176708}
}

@article{Appleby:2022itn,
doi = {10.3847/1538-4357/aca530},
url = {https://doi.org/10.3847/1538-4357/aca530},
year = {2023},
month = {jan},
publisher = {The American Astronomical Society},
volume = {942},
number = {2},
pages = {110},
author = {Appleby, Stephen and Kochappan, Joby P. and Chingangbam, Pravabati and Park, Changbom},
title = {Minkowski Tensors in Redshift Space—Beyond the Plane-parallel Approximation},
journal = {The Astrophysical Journal},}

@article{Collischon:2024jhw,
    author = {Collischon, Caroline and Klatt, Michael A. and Banday, Anthony J. and Sasaki, Manami and R\"ath, Christoph},
    title = "{Morphometry on the sphere: Cartesian and irreducible Minkowski tensors explained and implemented}",
    eprint = "2402.06286",
    archivePrefix = "arXiv",
    primaryClass = "astro-ph.IM",
    doi = "10.1038/s42005-024-01751-1",
    journal = "Commun. Phys.",
    volume = "7",
    number = "1",
    pages = "254",
    year = "2024"
}

@article{Kanafi:2023hmr,
    author = "Kanafi, M. H. Jalali and Movahed, S. M. S.",
    title = "{Probing the Anisotropy and Non-Gaussianity in the Redshift Space through the Conditional Moments of the First Derivative}",
    eprint = "2308.03086",
    archivePrefix = "arXiv",
    primaryClass = "astro-ph.CO",
    doi = "10.3847/1538-4357/ad1880",
    journal = "Astrophys. J.",
    volume = "963",
    number = "1",
    pages = "31",
    year = "2024"
}

@article{Kanafi:2025gpn,
    author = "Kanafi, M. H. Jalali and Movahed, S. M. S.",
    title = "{Quantifying weighted morphological content of large-scale structures via simulation-based inference}",
    eprint = "2511.03636",
    archivePrefix = "arXiv",
    primaryClass = "astro-ph.CO",
    doi = "10.1103/yw4d-vssh",
    journal = "Phys. Rev. D",
    volume = "113",
    number = "6",
    pages = "063543",
    year = "2026"
}

@article{Aghanim:2018eyx,
      author         = "Aghanim, N. and others",
      title          = "{Planck 2018 results. VI. Cosmological parameters}",
      collaboration  = "Planck",
      year           = "2018",
      eprint         = "1807.06209",
      archivePrefix  = "arXiv",
      primaryClass   = "astro-ph.CO",
      SLACcitation   = "%%CITATION = ARXIV:1807.06209;%%"
}

@article{McMullen:1997,
      author         = "P.~McMullen",
      title          = "{Isometry covariant valuations on convex bodies}",
      journal        = "Rend.~Circ.~Palermo",
      volume         = "50",
      year           = "1997",
      pages          = "259-271"
}

@INPROCEEDINGS{2002LNP...600..238B,
   author = {{Beisbart}, C. and {Dahlke}, R. and {Mecke}, K. and {Wagner}, H.
	},
    title = "{Vector- and Tensor-Valued Descriptors for Spatial Patterns}",
booktitle = {Morphology of Condensed Matter},
     year = 2002,
   series = {Lecture Notes in Physics, Berlin Springer Verlag},
   volume = 600,
    pages = {238-260},
}

@Article{HugSchSch07,
author = "Daniel Hug and Rolf Schneider and Ralph Schuster",
title="The space of isometry covariant tensor valuations",
journal="St. Petersburg Math. J.",
year="2008",
volume="19",
pages="137-158",
doi="10.1090/S1061-0022-07-00990-9",
}

@article{Kerscher:1998gs,
    author = "Kerscher, Martin and Schmalzing, Jens and Buchert, Thomas and Wagner, Herbert",
    title = "{Fluctuations in the IRAS 1.2 Jy catalog}",
    eprint = "astro-ph/9704028",
    archivePrefix = "arXiv",
    reportNumber = "SFB-375-180",
    journal = "Astron. Astrophys.",
    volume = "333",
    pages = "1--12",
    year = "1998"
}

@article{Kapahtia:2019ksk,
    author = "Kapahtia, Akanksha and Chingangbam, Pravabati and Appleby, Stephen",
    title = "{Morphology of 21cm brightness temperature during the Epoch of Reionization using Contour Minkowski Tensor}",
    eprint = "1904.06840",
    archivePrefix = "arXiv",
    primaryClass = "astro-ph.CO",
    doi = "10.1088/1475-7516/2019/09/053",
    journal = "JCAP",
    volume = "09",
    pages = "053",
    year = "2019"
}

@ARTICLE{Ganesan:2017,
       author = {{Ganesan}, Vidhya and {Chingangbam}, Pravabati},
        title = "{Tensor Minkowski Functionals: first application to the CMB}",
      journal = {\jcap},
         year = 2017,
        month = jun,
       volume = {2017},
       number = {6},
          eid = {023},
        pages = {023},
          doi = {10.1088/1475-7516/2017/06/023},
archivePrefix = {arXiv},
       eprint = {1608.07452},
 primaryClass = {astro-ph.CO}
}

@article{Chingangbam:2017uqv,
    author = "Chingangbam, Pravabati and Yogendran, K P and K., Joby P. and Ganesan, Vidhya and Appleby, Stephen and Park, Changbom",
    title = "{Tensor Minkowski Functionals for random fields on the sphere}",
    eprint = "1707.04386",
    archivePrefix = "arXiv",
    primaryClass = "astro-ph.CO",
    doi = "10.1088/1475-7516/2017/12/023",
    journal = "JCAP",
    volume = "12",
    pages = "023",
    year = "2017"
}

@ARTICLE{Chingangbam:2017PhLB,
       author = {{Chingangbam}, Pravabati and {Ganesan}, Vidhya and {Yogendran}, K.~P. and {Park}, Changbom},
        title = "{On Minkowski Functionals of CMB polarization}",
      journal = {Physics Letters B},
         year = 2017,
        month = aug,
       volume = {771},
        pages = {67-73},
          doi = {10.1016/j.physletb.2017.05.030},
archivePrefix = {arXiv},
       eprint = {1705.04454},
 primaryClass = {astro-ph.CO}
}

@ARTICLE{Chingangbam:2013,
       author = {{Chingangbam}, Pravabati and {Park}, Changbom},
        title = "{Residual foreground contamination in the WMAP data and bias in non-Gaussianity estimation}",
      journal = {\jcap},
         year = 2013,
        month = feb,
       volume = {2013},
       number = {2},
          eid = {031},
        pages = {031},
          doi = {10.1088/1475-7516/2013/02/031},
archivePrefix = {arXiv},
       eprint = {1210.2250},
 primaryClass = {astro-ph.CO}
}

@ARTICLE{Rahman:2021,
       author = {{Rahman}, Fazlu and {Chingangbam}, Pravabati and {Ghosh}, Tuhin},
        title = "{The nature of non-Gaussianity and statistical isotropy of the 408 MHz Haslam synchrotron map}",
      journal = {\jcap},
         year = 2021,
        month = jul,
       volume = {2021},
       number = {7},
          eid = {026},
        pages = {026},
          doi = {10.1088/1475-7516/2021/07/026},
archivePrefix = {arXiv},
       eprint = {2104.00419},
 primaryClass = {astro-ph.CO}
}

@ARTICLE{Chingangbam:2021,
       author = {{Chingangbam}, Pravabati and {Goyal}, Priya and {Yogendran}, K.~P. and {Appleby}, Stephen},
        title = "{Geometrical meaning of statistical isotropy of smooth random fields in two dimensions}",
      journal = {\prd},
         year = 2021,
        month = dec,
       volume = {104},
       number = {12},
          eid = {123516},
        pages = {123516},
          doi = {10.1103/PhysRevD.104.123516},
archivePrefix = {arXiv},
       eprint = {2109.05726},
 primaryClass = {astro-ph.CO}
}

@ARTICLE{Kapahtia:2018,
       author = {{Kapahtia}, Akanksha and {Chingangbam}, Pravabati and {Appleby}, Stephen and {Park}, Changbom},
        title = "{A novel probe of ionized bubble shape and size statistics of the epoch of reionization using the contour Minkowski Tensor}",
      journal = {\jcap},
         year = 2018,
        month = oct,
       volume = {2018},
       number = {10},
          eid = {011},
        pages = {011},
          doi = {10.1088/1475-7516/2018/10/011},
archivePrefix = {arXiv},
       eprint = {1712.09195},
 primaryClass = {astro-ph.CO}
}

@article{Bashir:2025gen,
    author = "Bashir, Masroor and S, Nidharssan and Chingangbam, Pravabati and Rahman, Fazlu and Goyal, Priya and Appleby, Stephen and Park, Changbom",
    title = "{Local patch analysis of ACT DR6 convergence map using morphological statistics}",
    eprint = "2503.17849",
    archivePrefix = "arXiv",
    primaryClass = "astro-ph.CO",
    month = "3",
    year = "2025"
}

@ARTICLE{Goyal:2019vkq,
       author = {{Goyal}, Priya and {Chingangbam}, Pravabati and {Appleby}, Stephen},
        title = "{Morphology of CMB fields{\textemdash}effect of weak gravitational lensing}",
      journal = {\jcap},
         year = 2020,
        month = feb,
       volume = {2020},
       number = {2},
          eid = {020},
        pages = {020},
          doi = {10.1088/1475-7516/2020/02/020},
archivePrefix = {arXiv},
       eprint = {1911.04740},
 primaryClass = {astro-ph.CO}
}

@article{Appleby:2017uvb,
      author         = "Appleby, Stephen and Chingangbam, Pravabati and Park,
                        Changbom and Hong, Sungwook E. and Kim, Juhan and Ganesan,
                        Vidhya",
      title          = "{Minkowski Tensors in Two Dimensions - Probing the
                        Morphology and Isotropy of the Matter and Galaxy Density
                        Fields}",
      journal        = "ApJ.",
      volume         = "858",
      year           = "2018",
      number         = "2",
      pages          = "87",
}

@article{K.:2018wpn,
      author         = "Joby, P. K. and Chingangbam, Pravabati and Ghosh, Tuhin
                        and Ganesan, Vidhya and Ravikumar, C. D.",
      title          = "{Search for anomalous alignments of structures in Planck
                        data using Minkowski Tensors}",
      journal        = "JCAP",
      volume         = "1901",
      year           = "2019",
      number         = "01",
      pages          = "009",
}

@article{Joby:2021,
  title = {Application of the contour Minkowski tensor and $\mathcal{D}$ statistic to the Planck $E$-mode data},
  author = {Joby, P. K. and Sen, Aparajita and Ghosh, Tuhin and Chingangbam, Pravabati and Basak, Soumen},
  journal = {Phys. Rev. D},
  volume = {103},
  issue = {12},
  pages = {123523},
  numpages = {10},
  year = {2021},
  month = {Jun},
  publisher = {American Physical Society},
  doi = {10.1103/PhysRevD.103.123523},
}

@ARTICLE{Rana:2018,
       author = {{Rana}, Sandeep and {Ghosh}, Tuhin and {Bagla}, Jasjeet S. and {Chingangbam}, Pravabati},
        title = "{Non-Gaussianity of diffuse Galactic synchrotron emission at 408 MHz}",
      journal = {\mnras},
         year = 2018,
        month = nov,
       volume = {481},
       number = {1},
        pages = {970-980},
          doi = {10.1093/mnras/sty2348},
archivePrefix = {arXiv},
       eprint = {1806.01565},
 primaryClass = {astro-ph.CO}
}

@article{Beisbart:2001vb,
      author         = "Beisbart, C. and Valdarnini, R. and Buchert, T.",
      title          = "{The morphological and dynamical evolution of simulated
                        galaxy clusters}",
      journal        = "Astron. Astrophys.",
      volume         = "379",
      year           = "2001",
      pages          = "412-425",
}

@article{Beisbart:2001gk,
      author         = "Beisbart, Claus and Buchert, Thomas and Wagner, Herbert",
      title          = "{Morphometry of spatial patterns}",
      journal        = "Physica",
      volume         = "A293",
      year           = "2001",
      pages          = "592-604",
}

@article{vandeWeygaert:2011hyr,
    author = "van de Weygaert, Rien and others",
    title = "{Alpha, Betti and the Megaparsec Universe: on the Topology of the Cosmic Web}",
    eprint = "1306.3640",
    archivePrefix = "arXiv",
    primaryClass = "astro-ph.CO",
    journal = "Trans. Comput. Sci.",
    volume = "14",
    pages = "60--101",
    year = "2011"
}

@article{Wilding:2020oza,
    author = "Wilding, Georg and Nevenzeel, Keimpe and van de Weygaert, Rien and Vegter, Gert and Pranav, Pratyush and Jones, Bernard J. T. and Efstathiou, Konstantinos and Feldbrugge, Job",
    title = "{Persistent homology of the cosmic web \textendash{} I. Hierarchical topology in \ensuremath{\Lambda}CDM cosmologies}",
    eprint = "2011.12851",
    archivePrefix = "arXiv",
    primaryClass = "astro-ph.CO",
    doi = "10.1093/mnras/stab2326",
    journal = "Mon. Not. Roy. Astron. Soc.",
    volume = "507",
    number = "2",
    pages = "2968--2990",
    year = "2021"
}

@ARTICLE{1989ApJ...345..618M,
   author = {{Melott}, A.~L. and {Cohen}, A.~P. and {Hamilton}, A.~J.~S. and 
	{Gott},    J.~R. and {Weinberg}, D.~H.},
    title = "{Topology of large-scale structure. IV - Topology in two dimensions}",
  journal        = "ApJ.",
     year = 1989,
    month = oct,
   volume = 345,
    pages = {618-626},
      doi = {10.1086/167935}
}

@ARTICLE{2001ApJ...553...33P,
   author = {{Park}, C. and {Gott},    J.~R. and {Choi}, Y.~J.},
    title = "{Topology of the Galaxy Distribution in the Hubble Deep Fields}",
  journal        = "ApJ.",
     year = 2001,
    month = may,
   volume = 553,
    pages = {33-38},
      doi = {10.1086/320640}
}

@article{10.1093/mnras/staf1110,
    author = {Afzal, Adeela and Alakhras, M and Kanafi, M H Jalali and Movahed, S M S},
    title = {Cosmic strings-induced CMB anisotropies in light of weighted morphology},
    journal = {Monthly Notices of the Royal Astronomical Society},
    volume = {541},
    number = {4},
    pages = {3851-3868},
    year = {2025},
    month = {07},
    issn = {0035-8711},
    doi = {10.1093/mnras/staf1110},
    eprint = {https://academic.oup.com/mnras/article-pdf/541/4/3851/63712501/staf1110.pdf},
}

@ARTICLE{1992ApJ...387....1P,
   author = {{Park}, C. and {Gott},    J.~R. and {Melott}, A.~L. and {Karachentsev}, I.~D.
	},
    title = "{The topology of large-scale structure. VI - Slices of the universe}",
  journal        = "ApJ.",
     year = 1992,
    month = mar,
   volume = 387,
    pages = {1-8},
}

@article{Schmalzing:1995qn,
      author         = "Schmalzing, Jens and Kerscher, Martin and Buchert,
                        Thomas",
      title          = "{Minkowski functionals in cosmology}",
      booktitle      = "{Dark matter in the universe. Proceedings, 132nd course
                        of the International School of Physics *Enrico Fermi*,
                        Varenna, Italy, July 25-August 4, 1995}",
      journal        = "Proc. Int. Sch. Phys. Fermi",
      volume         = "132",
      year           = "1996",
      pages          = "281-291",
}

@ARTICLE{2003ApJ...584....1M,
   author = {{Matsubara}, T.},
    title = "{Statistics of Smoothed Cosmic Fields in Perturbation Theory. I. Formulation and Useful Formulae in Second-Order Perturbation Theory}",
  journal        = "ApJ.",
     year = 2003,
    month = feb,
   volume = 584,
    pages = {1-33},
      doi = {10.1086/345521}
}

@article{Park:2009ja,
      author         = "Park, Changbom and Kim, Young-Rae",
      title          = "{Large-Scale Structure of the Universe as a Cosmic
                        Standard Ruler}",
      journal        = "ApJ.",
      volume         = "715",
      year           = "2010",
      pages          = "L185",
}

@article{Gott:1989yj,
      author         = "Gott, III, J. Richard and Park, Changbom and Juszkiewicz,
                        Roman and Bies, William E. and Bennett, David P. and
                        Bouchet, Francois R. and Stebbins, Albert",
      title          = "{Topology of microwave background fluctuations: Theory}",
      journal = "ApJ",
      year = "1990",
      month = "mar",
      volume = "352",
      pages = "1",      
      reportNumber   = "PUPT-1136",
      SLACcitation   = "%%CITATION = PUPT-1136;%%"
}

@article{Schmalzing:1997uc,
      author         = "Schmalzing, Jens and Gorski, Krzysztof M.",
      title          = "{Minkowski functionals used in the morphological analysis
                        of cosmic microwave background anisotropy maps}",
      journal        = "MNRAS",
      volume         = "297",
      year           = "1998",
      pages          = "355",
}

@ARTICLE{1991ApJ...378..457P,
   author = {{Park}, C. and {Gott},    J.~R.},
    title = "{Dynamical evolution of topology of large-scale structure}",
  journal        = "ApJ.",
     year = 1991,
    month = sep,
   volume = 378,
    pages = {457-460},
}

@Book{Hadwiger,
 author    = "Hadwiger, Hugo",
 title     = "Vorlesungen über Inhalt, Oberfläche und Isoperimetrie",
 publisher = "Springer",
 year      =  1957,
 address   = "Grundlehren der mathematischen Wissenschaften",
 edition   = ""
}

@article{Codis:2013exa,
      author         = "Codis, Sandrine and Pichon, Christophe and Pogosyan,
                        Dmitry and Bernardeau, Francis and Matsubara, Takahiko",
      title          = "{Non-Gaussian Minkowski functionals \& extrema counts in
                        redshift space}",
      journal        = "MNRAS",
      volume         = "435",
      year           = "2013",
      pages          = "531-564",
}

@article{doi:10.1111/j.1365-2966.2010.18015.x,
author = {Zunckel, Caroline and Gott, III ,J. Richard and Lunnan, Ragnhild},
title = {Using the topology of large-scale structure to constrain dark energy},
journal = {MNRAS},
volume = {412},
number = {2},
pages = {1401},
year = {2011},
}

@article{Feldbrugge:2019tal,
      author         = "Feldbrugge, Job and van Engelen, Matti and van de
                        Weygaert, Rien and Pranav, Pratyush and Vegter, Gert",
      title          = "{Stochastic Homology of Gaussian vs. non-Gaussian Random
                        Fields: Graphs towards Betti Numbers and Persistence
                        Diagrams}",
      journal        = "JCAP",
      volume         = "1909",
      year           = "2019",
      number         = "09",
      pages          = "052",
      doi            = "10.1088/1475-7516/2019/09/052",
      eprint         = "1908.01619",
      archivePrefix  = "arXiv",
      primaryClass   = "astro-ph.CO",
      SLACcitation   = "%%CITATION = ARXIV:1908.01619;%%"
}

@article{Pranav:2018lox,
      author         = "Pranav, Pratyush and Adler, Robert J. and Buchert, Thomas
                        and Edelsbrunner, Herbert and Jones, Bernard J. T. and
                        Schwartzman, Armin and Wagner, Hubert and van de Weygaert,
                        Rien",
      title          = "{Unexpected Topology of the Temperature Fluctuations in
                        the Cosmic Microwave Background}",
      journal        = "Astron. Astrophys.",
      volume         = "627",
      year           = "2019",
      pages          = "A163",
      doi            = "10.1051/0004-6361/201834916",
      eprint         = "1812.07678",
      archivePrefix  = "arXiv",
      primaryClass   = "astro-ph.CO",
      SLACcitation   = "%%CITATION = ARXIV:1812.07678;%%"
}

@article{Pranav:2018pnu,
      author         = "Pranav, Pratyush and van de Weygaert, Rien and Vegter,
                        Gert and Jones, Bernard J. T. and Adler, Robert J. and
                        Feldbrugge, Job and Park, Changbom and Buchert, Thomas and
                        Kerber, Michael",
      title          = "{Topology and Geometry of Gaussian random fields I: on
                        Betti Numbers, Euler characteristic and Minkowski
                        functionals}",
      journal        = "Mon. Not. Roy. Astron. Soc.",
      volume         = "485",
      year           = "2019",
      number         = "3",
      pages          = "4167-4208",
      doi            = "10.1093/mnras/stz541",
      eprint         = "1812.07310",
      archivePrefix  = "arXiv",
      primaryClass   = "astro-ph.CO",
      SLACcitation   = "%%CITATION = ARXIV:1812.07310;%%"
}

@article{Pranav:2016gwr,
      author         = "Pranav, Pratyush and Edelsbrunner, Herbert and van de
                        Weygaert, Rien and Vegter, Gert and Kerber, Michael and
                        Jones, Bernard J. T. and Wintraecken, Mathijs",
      title          = "{The Topology of the Cosmic Web in Terms of Persistent
                        Betti Numbers}",
      journal        = "Mon. Not. Roy. Astron. Soc.",
      volume         = "465",
      year           = "2017",
      number         = "4",
      pages          = "4281-4310",
      doi            = "10.1093/mnras/stw2862",
      eprint         = "1608.04519",
      archivePrefix  = "arXiv",
      primaryClass   = "astro-ph.CO",
      SLACcitation   = "%%CITATION = ARXIV:1608.04519;%%"
}

@article{Shivshankar:2015aza,
      author         = "Shivshankar, Nithin and Pranav, Pratyush and Natarajan,
                        Vijay and van de Weygaert, Rien and Bos, E. G. Patrick and
                        Rieder, Steven",
      title          = "{Felix: A Topology based Framework for Visual Exploration
                        of Cosmic Filaments}",
      journal        = "Comput. Graphics",
      volume         = "1",
      year           = "2015",
      pages          = "1",
      doi            = "10.1109/TVCG.2015.2452919",
      eprint         = "1508.00737",
      archivePrefix  = "arXiv",
      primaryClass   = "astro-ph.CO",
      SLACcitation   = "%%CITATION = ARXIV:1508.00737;%%"
}

@article{vandeWeygaert:2011ip,
      author         = "van de Weygaert, Rien and others",
      title          = "{Probing Dark Energy with Alpha Shapes and Betti
                        Numbers}",
      year           = "2011",
      eprint         = "1110.5528",
      archivePrefix  = "arXiv",
      primaryClass   = "astro-ph.CO",
      SLACcitation   = "%%CITATION = ARXIV:1110.5528;%%"
}

@article{Schmalzing:1997aj,
      author         = "Schmalzing, Jens and Buchert, Thomas",
      title          = "{Beyond genus statistics: A Unifying approach to the
                        morphology of cosmic structure}",
      journal        = "Astrophys. J.",
      volume         = "482",
      year           = "1997",
      pages          = "L1-L4",
      doi            = "10.1086/310680",
      eprint         = "astro-ph/9702130",
      archivePrefix  = "arXiv",
      primaryClass   = "astro-ph",
      SLACcitation   = "%%CITATION = ASTRO-PH/9702130;%%"
}

@article{1367-2630-15-8-083028,
  author={G E Schroder-Turk and W Mickel and S C Kapfer and F M Schaller and B Breidenbach and D Hug and K Mecke},
  title={Minkowski tensors of anisotropic spatial structure},
  journal={New Journal of Physics},
  volume={15},
  number={8},
  pages={083028},
  year={2013},
}

@article{PhysRevE.77.051805,
  title = {Characterization of the dynamics of block copolymer microdomains with local morphological measures},
  author = {Rehse, S. and Mecke, K. and Magerle, R.},
  journal = {Phys. Rev. E},
  volume = {77},
  issue = {5},
  pages = {051805},
  numpages = {8},
  year = {2008},
  month = {May},
}

@article{Becker2003ComplexDS,
  title={Complex dewetting scenarios captured by thin-film models},
  author={J{\"u}rgen C Becker and G{\"u}nther Gr{\"u}n and Ralf Seemann and Hubert Mantz and Karin Jacobs and Klaus R. Mecke and Ralf Blossey},
  journal={Nature Materials},
  year={2003},
  volume={2},
  pages={59-63}
}

@article{Olszowka2006,
author = {Olszowka, Violetta and Hund, Markus and Kuntermann, Volker and Scherdel, Sabine and Tsarkova, Larisa and Boker, Alexander and Krausch, Georg},
year = {2006},
month = {12},
pages = {1089-1094},
title = {Large scale alignment of a lamellar block copolymer thin film via electric fields: A time-resolved SFM study},
volume = {2},
journal = {Soft Matter},
}

@Book{ nla.cat-vn1821482,
author = { Matheron, G. },
title = { Random sets and integral geometry },
publisher = { Wiley New York },
pages = { xxiii, 261 p. },
year = { 1974 },
type = { Book },
language = { English },
subjects = { Geometric probabilities.; Random sets.; Integral geometry.; Set theory. },
life-dates = { 1974 - 1975 },
}

@Book{nla.cat-vn2176896,
author = { Santalo, Luis A. },
title = { Integral geometry and geometric probability},
publisher = { Addison-Wesley Pub. Co., Advanced Book Program Reading, Mass },
pages = { xvii, 404 p. : },
year = { 1976 },
type = { Book },
language = { English },
subjects = { Geometric probabilities.; Geometry, Integral. },
life-dates = { 1976 -  },
catalogue-url = { https://nla.gov.au/nla.cat-vn2176896 },
}

@article {JMI:JMI3331,
author = {Schroder-Turk, G.E. and Kapfer, S. and Breidenbach, B. and Beisbart, C. and Mecke, K.},
title = {Tensorial Minkowski functionals and anisotropy measures for planar patterns},
journal = {Journal of Microscopy},
volume = {238},
number = {1},
publisher = {Blackwell Publishing Ltd},
issn = {1365-2818},
pages = {57--74},
year = {2010},
}

@article{Sahni:1998cr,
    author = "Sahni, Varun and Sathyaprakash, B. S. and Shandarin, Sergei F.",
    title = "{Shapefinders: A New shape diagnostic for large scale structure}",
    eprint = "astro-ph/9801053",
    archivePrefix = "arXiv",
    reportNumber = "IUCAA-2-98",
    doi = "10.1086/311214",
    journal = "Astrophys. J. Lett.",
    volume = "495",
    pages = "L5--L8",
    year = "1998"
}

@article{Bharadwaj:1999jm,
    author = "Bharadwaj, Somnath and Sahni, Varun and Sathyaprakash, B. S. and Shandarin, Sergei F. and Yess, Capp",
    title = "{Evidence for filamentarity in the las campanas redshift survey}",
    eprint = "astro-ph/9904406",
    archivePrefix = "arXiv",
    reportNumber = "IUCAA-18-99",
    doi = "10.1086/308163",
    journal = "Astrophys. J.",
    volume = "528",
    pages = "21",
    year = "2000"
}

@article{Klatt_2022,
	year = 2022,
	month = {apr},
	publisher = {{IOP} Publishing},
	volume = {2022},
	number = {4},
	pages = {043301},
	author = {Klatt, Michael Andreas and H{\"o}rmann, Max and Mecke, Klaus},
	title = {Characterization of anisotropic Gaussian random fields by Minkowski tensors},
	journal = {Journal of Statistical Mechanics: Theory and Experiment},
}

@article{Schaller2020,
  doi = {10.21105/joss.02538},
  year = {2020},
  publisher = {The Open Journal},
  volume = {5},
  number = {54},
  pages = {2538},
  author = {Fabian M. Schaller and Jenny Wagner and Sebastian C. Kapfer},
  title = {papaya2: 2D Irreducible Minkowski Tensor computation},
  journal = {Journal of Open Source Software}
}

@article{Munshi:2020tzm,
    author = "Munshi, D. and Namikawa, T. and McEwen, J. D. and Kitching, T. D. and Bouchet, F. R.",
    title = "{Morphology of weak lensing convergence maps}",
    eprint = "2010.05669",
    archivePrefix = "arXiv",
    primaryClass = "astro-ph.CO",
    doi = "10.1093/mnras/stab2101",
    journal = "Mon. Not. Roy. Astron. Soc.",
    volume = "507",
    number = "1",
    pages = "1421--1433",
    year = "2021"
}

@article{Appleby:2021xoz,
    author = "Appleby, Stephen and Park, Changbom and Pranav, Pratyush and Hong, Sungwook E. and Hwang, Ho Seong and Kim, Juhan and Buchert, Thomas",
    title = "{Minkowski Functionals of SDSS-III BOSS: Hints of Possible Anisotropy in the Density Field?}",
    eprint = "2110.06109",
    archivePrefix = "arXiv",
    primaryClass = "astro-ph.CO",
    doi = "10.3847/1538-4357/ac562a",
    journal = "Astrophys. J.",
    volume = "928",
    number = "2",
    pages = "108",
    year = "2022"
}

@ARTICLE{Kapahtia:2021,
       author = {{Kapahtia}, Akanksha and {Chingangbam}, Pravabati and {Ghara}, Raghunath and {Appleby}, Stephen and {Choudhury}, Tirthankar Roy},
        title = "{Prospects of constraining reionization model parameters using Minkowski tensors and Betti numbers}",
      journal = {\jcap},
         year = 2021,
        month = may,
       volume = {2021},
       number = {5},
          eid = {026},
        pages = {026},
          doi = {10.1088/1475-7516/2021/05/026},
archivePrefix = {arXiv},
       eprint = {2101.03962},
 primaryClass = {astro-ph.CO}
}

@article{Liu:2025haj,
    author = "Liu, Wei and Paillas, Enrique and Cuesta-Lazaro, Carolina and Valogiannis, Georgios and Fang, Wenjuan",
    title = "{Cosmological constraints from the Minkowski functionals of the BOSS CMASS galaxy sample}",
    eprint = "2501.01698",
    archivePrefix = "arXiv",
    primaryClass = "astro-ph.CO",
    doi = "10.1088/1475-7516/2025/05/064",
    journal = "JCAP",
    volume = "05",
    pages = "064",
    year = "2025"
}

@article{Liu:2022vtr,
    author = "Liu, Wei and Jiang, Aoxiang and Fang, Wenjuan",
    title = "{Probing massive neutrinos with the Minkowski functionals of large-scale structure}",
    eprint = "2204.02945",
    archivePrefix = "arXiv",
    primaryClass = "astro-ph.CO",
    doi = "10.1088/1475-7516/2022/07/045",
    journal = "JCAP",
    volume = "07",
    number = "07",
    pages = "045",
    year = "2022"
}

@article{Matsubara:2022ohx,
    author = "Matsubara, Takahiko",
    title = "{Integrated perturbation theory for cosmological tensor fields. I. Basic formulation}",
    eprint = "2210.10435",
    archivePrefix = "arXiv",
    primaryClass = "astro-ph.CO",
    reportNumber = "KEK-Cosmo-0299, KEK-TH-2465",
    doi = "10.1103/PhysRevD.110.063543",
    journal = "Phys. Rev. D",
    volume = "110",
    number = "6",
    pages = "063543",
    year = "2024"
}

@article{Matsubara:2022eui,
    author = "Matsubara, Takahiko",
    title = "{Integrated perturbation theory for cosmological tensor fields. II. Loop corrections}",
    eprint = "2210.11085",
    archivePrefix = "arXiv",
    primaryClass = "astro-ph.CO",
    reportNumber = "KEK-Cosmo-0300, KEK-TH-2466",
    doi = "10.1103/PhysRevD.110.063544",
    journal = "Phys. Rev. D",
    volume = "110",
    number = "6",
    pages = "063544",
    year = "2024"
}

@article{Matsubara:2023avg,
    author = "Matsubara, Takahiko",
    title = "{Integrated perturbation theory for cosmological tensor fields. III. Projection effects}",
    eprint = "2304.13304",
    archivePrefix = "arXiv",
    primaryClass = "astro-ph.CO",
    reportNumber = "KEK-TH-2520, KEK-Cosmo-0312",
    doi = "10.1103/PhysRevD.110.063545",
    journal = "Phys. Rev. D",
    volume = "110",
    number = "6",
    pages = "063545",
    year = "2024"
}

@article{Matsubara:2024sqn,
    author = "Matsubara, Takahiko",
    title = "{Integrated perturbation theory for cosmological tensor fields. IV. Full-sky formulation}",
    eprint = "2405.09038",
    archivePrefix = "arXiv",
    primaryClass = "astro-ph.CO",
    reportNumber = "KEK-Cosmo-0343",
    doi = "10.1103/PhysRevD.110.063546",
    journal = "Phys. Rev. D",
    volume = "110",
    number = "6",
    pages = "063546",
    year = "2024"
}

@article{Abedi:2024ajq,
    author = "Abedi, Fatemeh and Kanafi, Mohammad Hossein Jalali and Movahed, Seyed Mohammad Sadegh",
    title = "{Impact of redshift space distortions on persistent homology of cosmic matter density field}",
    eprint = "2410.01751",
    archivePrefix = "arXiv",
    primaryClass = "astro-ph.CO",
    doi = "10.1142/S0219887825400067",
    journal = "Int. J. Geom. Meth. Mod. Phys.",
    volume = "22",
    number = "10",
    pages = "2540006",
    year = "2025"
}

@article{Liu:2023qrj,
    author = "Liu, Wei and Jiang, Aoxiang and Fang, Wenjuan",
    title = "{Probing massive neutrinos with the Minkowski functionals of the galaxy distribution}",
    eprint = "2302.08162",
    archivePrefix = "arXiv",
    primaryClass = "astro-ph.CO",
    doi = "10.1088/1475-7516/2023/09/037",
    journal = "JCAP",
    volume = "09",
    pages = "037",
    year = "2023"
}

@Article{Alesker1999,
author={Alesker, S.},
title={Description of Continuous Isometry Covariant Valuations on Convex Sets},
journal={Geometriae Dedicata},
year={1999},
month={Mar},
day={01},
volume={74},
number={3},
pages={241-248},
issn={1572-9168},
doi={10.1023/A:1005035232264},
}

@ARTICLE{2020arXiv200615038K,
       author = {{Kochappan}, Joby P. and {Sen}, Aparajita and {Ghosh}, Tuhin and
                 {Chingangbam}, Pravabati and {Basak}, Soumen},
        title = "{Statistical Isotropy of the CMB E-mode signal}",
         year = 2020,
archivePrefix = {arXiv},
       eprint = {2006.15038},
 primaryClass = {astro-ph.CO},
          doi = {10.48550/arXiv.2006.15038}
}

@ARTICLE{2019ApJ...887..128A,
       author = {{Appleby}, Stephen and {Kochappan}, Joby P. and {Chingangbam}, Pravabati and {Park}, Changbom},
        title = "{Ensemble Average of Three-dimensional Minkowski Tensors of a Gaussian Random Field in Redshift Space}",
      journal = {\apj},
         year = 2019,
        month = dec,
       volume = {887},
       number = {2},
          eid = {128},
        pages = {128},
          doi = {10.3847/1538-4357/ab5057},
archivePrefix = {arXiv},
       eprint = {1908.02440},
 primaryClass = {astro-ph.CO}
}

@ARTICLE{1994A&A...288..697M,
       author = {{Mecke}, K.~R. and {Buchert}, T. and {Wagner}, H.},
        title = "{Robust morphological measures for large-scale structure in the Universe}",
      journal = {\aap},
         year = 1994,
        month = aug,
       volume = {288},
        pages = {697-704},
          doi = {10.48550/arXiv.astro-ph/9312028},
archivePrefix = {arXiv},
       eprint = {astro-ph/9312028},
 primaryClass = {astro-ph}
}

@article{Park:2013dga,
    author = "Park, Changbom and Pranav, Pratyush and Chingangbam, Pravabati and van de Weygaert, Rien and Jones, Bernard and Vegter, Gert and Kim, Inkang and Hidding, Johan and Hellwing, Wojciech A.",
    title = "{Betti numbers of Gaussian fields}",
    eprint = "1307.2384",
    archivePrefix = "arXiv",
    primaryClass = "astro-ph.CO",
    doi = "10.5303/JKAS.2013.46.3.125",
    journal = "J. Korean Astron. Soc.",
    volume = "46",
    number = "3",
    pages = "125--131",
    year = "2013"
}

@article{https://doi.org/10.1111/j.1365-2966.2008.13674.x,
author = {Hikage, C. and Matsubara, T. and Coles, P. and Liguori, M. and Hansen, F. K. and Matarrese, S.},
title = {Limits on primordial non-Gaussianity from Minkowski Functionals of the WMAP temperature anisotropies},
journal = {Monthly Notices of the Royal Astronomical Society},
volume = {389},
number = {3},
pages = {1439-1446},
doi = {10.1111/j.1365-2966.2008.13674.x},
    eprint = "0802.3677",
year = {2008}
}

@ARTICLE{2016MNRAS.461.1363N,
       author = {{Novaes}, C.~P. and {Bernui}, A. and {Marques}, G.~A. and {Ferreira}, I.~S.},
        title = "{Local analyses of Planck maps with Minkowski functionals}",
      journal = {\mnras},
         year = 2016,
        month = sep,
       volume = {461},
       number = {2},
        pages = {1363-1373},
          doi = {10.1093/mnras/stw1427},
archivePrefix = {arXiv},
       eprint = {1606.04075},
 primaryClass = {astro-ph.CO}
}

@article{Jones:2023ncn,
    author = "Jones, Joann and Copi, Craig J. and Starkman, Glenn D. and Akrami, Yashar",
    title = "{Strong Evidence Against a Statistically Isotropic Universe}",
    eprint = "2310.12859",
    archivePrefix = "arXiv",
    primaryClass = "astro-ph.CO",
    month = "10",
    year = "2023"
}

@article{Secrest:2020has,
    author = "Secrest, Nathan J. and von Hausegger, Sebastian and Rameez, Mohamed and Mohayaee, Roya and Sarkar, Subir and Colin, Jacques",
    title = "{A Test of the Cosmological Principle with Quasars}",
    eprint = "2009.14826",
    archivePrefix = "arXiv",
    primaryClass = "astro-ph.CO",
    doi = "10.3847/2041-8213/abdd40",
    journal = "Astrophys. J. Lett.",
    volume = "908",
    number = "2",
    pages = "L51",
    year = "2021"
}

@article{Battye:2026tsu,
    author = "Battye, Richard A. and Moss, Adam",
    title = "{How isotropic is dark energy?}",
    eprint = "2601.22351",
    archivePrefix = "arXiv",
    primaryClass = "astro-ph.CO",
    month = "1",
    year = "2026"
}

@article{Goyal:2021nun,
    author = "Goyal, Priya and Chingangbam, Pravabati",
    title = "{Local patch analysis for testing statistical isotropy of the Planck convergence map}",
    eprint = "2104.00418",
    archivePrefix = "arXiv",
    primaryClass = "astro-ph.CO",
    doi = "10.1088/1475-7516/2021/08/006",
    journal = "JCAP",
    volume = "08",
    pages = "006",
    year = "2021"
}

@article{CarronDuque:2026unb,
    author = "Carr{\'o}n Duque, Javier and Martin Barandiaran, Mikel and Mart{\'\i}nez-Arrizabalaga, Joseba",
    title = "{More Than Power: Revisiting the CMB Hemispherical Power Asymmetry with Morphological Descriptors}",
    eprint = "2603.22449",
    archivePrefix = "arXiv",
    primaryClass = "astro-ph.CO",
    reportNumber = "IFT-UAM/CSIC-26-036",
    month = "3",
    year = "2026"
}

@article{Eriksen:2003db,
    author = "Eriksen, H. K. and Hansen, F. K. and Banday, A. J. and Gorski, K. M. and Lilje, P. B.",
    title = "{Asymmetries in the Cosmic Microwave Background anisotropy field}",
    eprint = "astro-ph/0307507",
    archivePrefix = "arXiv",
    doi = "10.1086/382267",
    journal = "Astrophys. J.",
    volume = "605",
    pages = "14--20",
    year = "2004",
    note = "[Erratum: Astrophys.J. 609, 1198 (2004)]"
}

@article{Hansen:2008ym,
    author = "Hansen, F. K. and Banday, A. J. and Gorski, K. M. and Eriksen, H. K. and Lilje, P. B.",
    title = "{Power Asymmetry in Cosmic Microwave Background Fluctuations from Full Sky to Sub-degree Scales: Is the Universe Isotropic?}",
    eprint = "0812.3795",
    archivePrefix = "arXiv",
    primaryClass = "astro-ph",
    doi = "10.1088/0004-637X/704/2/1448",
    journal = "Astrophys. J.",
    volume = "704",
    pages = "1448--1458",
    year = "2009"
}

@ARTICLE{Rubart:2013,
       author = {{Rubart}, M. and {Schwarz}, D.~J.},
        title = "{Cosmic radio dipole from NVSS and WENSS}",
      journal = {\aap},
         year = 2013,
        month = jul,
       volume = {555},
          eid = {A117},
        pages = {A117},
          doi = {10.1051/0004-6361/201321215},
archivePrefix = {arXiv},
       eprint = {1301.5559},
 primaryClass = {astro-ph.CO},
       adsurl = {https://ui.adsabs.harvard.edu/abs/2013A&A...555A.117R}
}

@ARTICLE{Singal:2019,
       author = {{Singal}, Ashok K.},
        title = "{Large disparity in cosmic reference frames determined from the sky distributions of radio sources and the microwave background radiation}",
      journal = {\prd},
         year = 2019,
        month = sep,
       volume = {100},
       number = {6},
          eid = {063501},
        pages = {063501},
          doi = {10.1103/PhysRevD.100.063501},
archivePrefix = {arXiv},
       eprint = {1904.11362},
 primaryClass = {physics.gen-ph},
       adsurl = {https://ui.adsabs.harvard.edu/abs/2019PhRvD.100f3501S}
}

@article{Hajian:2003qq,
    author = "Hajian, Amir and Souradeep, Tarun",
    title = "{Measuring statistical isotropy of the CMB anisotropy}",
    eprint = "astro-ph/0308001",
    archivePrefix = "arXiv",
    reportNumber = "IUCAA-37-2003",
    doi = "10.1086/379757",
    journal = "Astrophys. J. Lett.",
    volume = "597",
    pages = "L5--L8",
    year = "2003"
}

@article{Hajian:2005jh,
    author = "Hajian, Amir and Souradeep, Tarun",
    title = "{The Cosmic microwave background bipolar power spectrum: Basic formalism and applications}",
    eprint = "astro-ph/0501001",
    archivePrefix = "arXiv",
    month = "1",
    year = "2005"
}

@article{Minato:2025ozy,
    author = "Minato, Keita and Taruya, Atsushi and Okumura, Teppei and Shiraishi, Maresuke",
    title = "{Probing dipolar power asymmetry with galaxy clustering and intrinsic alignments}",
    eprint = "2505.19941",
    archivePrefix = "arXiv",
    primaryClass = "astro-ph.CO",
    reportNumber = "KUNS-3054, YITP-25-82",
    month = "5",
    year = "2025"
}

@article{Inoue:2024twk,
    author = "Inoue, Takuya and Okumura, Teppei and Saga, Shohei and Taruya, Atsushi",
    title = "{Information content in anisotropic cosmological fields: Impact of different multipole expansion scheme for galaxy density and ellipticity correlations}",
    eprint = "2406.19669",
    archivePrefix = "arXiv",
    primaryClass = "astro-ph.CO",
    reportNumber = "YITP-24-77",
    doi = "10.1103/PhysRevD.111.023507",
    journal = "Phys. Rev. D",
    volume = "111",
    number = "2",
    pages = "023507",
    year = "2025"
}

@ARTICLE{1991ApJ...380....1M,
       author = {{Miralda-Escude}, Jordi},
        title = "{The Correlation Function of Galaxy Ellipticities Produced by Gravitational Lensing}",
      journal = {\apj},
         year = 1991,
        month = oct,
       volume = {380},
        pages = {1},
          doi = {10.1086/170555},
       adsurl = {https://ui.adsabs.harvard.edu/abs/1991ApJ...380....1M}
}

@article{Kaiser:1991qi,
    author = "Kaiser, Nick",
    title = "{Weak gravitational lensing of distant galaxies}",
    reportNumber = "CITA-91-14",
    doi = "10.1086/171151",
    journal = "Astrophys. J.",
    volume = "388",
    pages = "272",
    year = "1992"
}

@article{Schneider:2002jd,
    author = "Schneider, Peter and van Waerbeke, Ludovic and Kilbinger, Martin and Mellier, Yannick",
    title = "{Analysis of two-point statistics of cosmic shear: I. estimators and covariances}",
    eprint = "astro-ph/0206182",
    archivePrefix = "arXiv",
    doi = "10.1051/0004-6361:20021341",
    journal = "Astron. Astrophys.",
    volume = "396",
    pages = "1--20",
    year = "2002"
}

@article{Bartelmann:1999yn,
    author = "Bartelmann, Matthias and Schneider, Peter",
    title = "{Weak gravitational lensing}",
    eprint = "astro-ph/9912508",
    archivePrefix = "arXiv",
    doi = "10.1016/S0370-1573(00)00082-X",
    journal = "Phys. Rept.",
    volume = "340",
    pages = "291--472",
    year = "2001"
}

@article{Appleby:2025sbk,
    author = "Appleby, Stephen and Pichon, Christophe and Chingangbam, Pravabati and Pogosyan, Dmitri and Park, Changbom",
    title = "{Non-Gaussian Expansion of Minkowski Tensors in Redshift Space}",
    eprint = "2507.10091",
    archivePrefix = "arXiv",
    primaryClass = "astro-ph.CO",
    doi = "10.3847/1538-4357/ae3a8b",
    journal = "Astrophys. J.",
    volume = "1001",
    number = "2",
    pages = "213",
    year = "2026"
}

@article{Aluri:2022hzs,
    author = "Aluri, Pavan Kumar and others",
    title = "{Is the observable Universe consistent with the cosmological principle?}",
    eprint = "2207.05765",
    archivePrefix = "arXiv",
    primaryClass = "astro-ph.CO",
    doi = "10.1088/1361-6382/acbefc",
    journal = "Class. Quant. Grav.",
    volume = "40",
    number = "9",
    pages = "094001",
    year = "2023"
}
\bibliographystyle{JHEP}

\end{document}